\documentclass[12pt]{article}
\usepackage{authblk}
\usepackage{natbib}
\usepackage [autostyle, english = american]{csquotes}
\usepackage[colorlinks=true,linkcolor=blue,citecolor=blue]{hyperref}%
\usepackage[absolute,overlay]{textpos}
\usepackage[utf8]{inputenc}
\usepackage[usenames,dvipsnames]{color}
\usepackage{comment}
\usepackage{bookmark}
\usepackage{booktabs}
\usepackage{enumitem}
\usepackage{blindtext}
\usepackage[center]{caption}
\usepackage{multirow}
\usepackage{amsmath}
\usepackage[labelfont=bf]{caption}
\usepackage{mathtools}
\usepackage{amsfonts}
\usepackage{comment}
\usepackage{amssymb}
\usepackage{graphicx}
\usepackage{indentfirst}
\usepackage{titling}
\usepackage{tikz}
\usepackage{qtree}
\usetikzlibrary{snakes}
\usetikzlibrary{mindmap}
\usepackage{lscape}
\usepackage{rotating}
\usepackage{setspace}
\usepackage[capposition=top]{floatrow}
\usepackage{lineno}
\usepackage{boldline, cellspace, makecell, tabularx}
\usepackage{booktabs, siunitx}
\usepackage{subfigure}
\usepackage{color,hyperref}
\usepackage{hyperref}
\usepackage{fancyhdr,titlesec,lipsum}
\newcommand\cites[1]{\citeauthor{#1}'s\ (\citeyear{#1})}

\def\sym#1{\ifmmode^{#1}\else\(^{#1}\)\fi}
\usetikzlibrary{arrows,calc,positioning}
\usepackage{chngcntr}
\usepackage{appendix}
\usepackage{scrwfile}
\usepackage{soul,color}
\soulregister\cite7
\soulregister\ref7
\soulregister\pageref7

\usepackage{mfirstuc}
\MFUnocap{a}
\MFUnocap{for}
\MFUnocap{the}
\MFUnocap{of}

\title{\vspace*{-2cm}
\Huge{Location Matters: Insights from a Natural Field Experiment to Enhance Small Business Tax Compliance in Indonesia}}

\makeatletter
\renewcommand{\author}[1]{\gdef\@author{\par#1}}
\makeatother
\author{
	Sarah Xue Dong \\[-1mm]
	\small{ANU} \\ 
	Agung Satyadini \\[-1mm]
	\small{ANU} \\ 
	Mathias Sinning \\[-1mm]  
	\small{ANU, RWI, IZA, J-PAL}
}

\date{\today}

\begin{document}

\thispagestyle{empty}

\maketitle		

\vspace{-.5cm}
\noindent \textbf{Abstract.} Tax compliance among small businesses remains low in developing countries, yet little is known about how regional context shapes the effectiveness of enforcement strategies. Both theory and evidence suggest an ambiguous relationship between compliance and geographic proximity to tax offices. We study this issue using a large-scale natural field experiment with Indonesia’s tax authority involving~12,000 micro, small, and medium enterprises (MSMEs). Businesses were randomly assigned to receive deterrence, information, or public goods letters, or no message. All letters improved compliance, with deterrence messages producing the largest gains~-- substantially increasing filing rates and raising monthly tax payments. Each dollar spent on deterrence letters generated about~US\$30 in additional revenue over the course of a year. We observe high compliance among non-treated MSMEs near metropolitan tax offices and find that enforcement messages successfully raise compliance in non-metropolitan regions to comparable levels. However, targeting already compliant MSMEs near metropolitan tax offices backfires, underscoring the need for geographically tailored tax administration strategies. These results provide novel experimental evidence on the relation between geographic proximity and the effectiveness of tax enforcement, helping to reconcile mixed findings in the tax compliance literature. \\

\noindent \textbf{JEL-Classification:} C93, D90, H25, H26 \newline


\vfill

\noindent \rule{3cm}{0.01cm}

\small \renewcommand{\baselinestretch}{1}
\noindent {\footnotesize We thank Shawn Chen, Arya Gaduh, Yothin Jinjarak, Muhammad Ali Nasir, Donghyun Park, Maathumai Ranjan, Tanisa Tawichsri and Yon Arsal for helpful comments and gratefully acknowledge the financial support from the Indonesia Endowment Fund for Education (LPDP, S-3583). All correspondence to Sarah Xue Dong, Crawford School of Public Policy, JG Crawford Building \#132, Lennox Crossing, The Australian National University, Canberra ACT 2601, Tel: +61 2 6125 8282, E-mail: sarah.dong@anu.edu.au.}

\thispagestyle{empty}
\small \renewcommand{\baselinestretch}{1.5}
\normalsize 
\newpage

\section{Introduction}

\setcounter{page}{1}

\noindent Micro, small, and medium enterprises (MSMEs) play a vital role in the economic landscape of developing countries, driving employment and growth. In Indonesia, MSMEs account for about~99\% of businesses and generate around 62\% of GDP \citep{coordinating_ministry_for_economc_affairs_encourage_2022}. Yet, despite their economic significance, persistently low tax compliance leaves much of their fiscal potential untapped \citep{dgt_taxpayers_2021}, a missed opportunity to finance Indonesia’s development agenda \citep{bank_innovations_2022}. The challenge is compounded by structural constraints: the vast majority of small businesses remain unregistered, compliance among registered firms is low \citep{dgt_taxpayers_2021}, and the absence of third-party income reporting hampers the government’s ability to verify income and assess liabilities accurately. Under such conditions, it is crucial for policymakers to identify the realistic levels of compliance that can be achieved and design targeted strategies to enhance MSME tax compliance.

This paper explores the effectiveness of behavioral interventions designed to improve MSME tax compliance, using data from a Natural Field Experiment conducted in collaboration with Indonesia's Directorate General of Taxes (DGT). 
Our sample consists of 12,000 DGT-registered businesses randomly assigned to receive one of three hard-copy letters, each informed by insights from the behavioral literature: a deterrence letter, a tax education letter, and a letter emphasizing the use of tax revenue. A control group received no correspondence. The experiment was implemented across Jakarta (metropolitan capital), Palembang in South Sumatera (urban/semi-urban), and Mataram in West Nusa Tenggara (urban, semi-urban, and rural), covering~13 local tax office jurisdictions. We observe the filing and payment behavior of MSMEs over a 12-month period after the intervention. 

Our sample includes only registered businesses, allowing us to study variation among those already at the higher end of the compliance spectrum. It further allows us to examine regional variation in baseline compliance and treatment effectiveness. A key feature distinguishing the three study regions is the geographic proximity of businesses to their nearest tax office. For example, in Jakarta, half of the businesses in our sample are located within one kilometer of a tax office, with six such offices operating within a~147-square-kilometer area. In contrast, in West Nusa Tenggara, 90\% of sampled businesses are more than~23 kilometers away from the nearest tax office. To assess how spatial variation shapes outcomes, we conduct a heterogeneity analysis stratifying businesses by distance to the nearest tax office. This allows us to assess whether proximity of small businesses to tax offices moderates or amplifies the impact of behavioral interventions on compliance behavior. 

We provide the first experimental evidence on how geographic proximity of businesses to tax offices influences compliance. Proximity facilitates greater availability and efficient dissemination of information, which flows more readily near its source \citep{ivkovic_tax-motivated_2005}. As a consequence, businesses located in close proximity to a tax office may be less compliant if they perceive their proximity as providing an informational advantage \citep{kubick_irs_2017}. Conversely, it is also possible that tax offices collect more information about nearby businesses than those businesses obtain about tax offices \citep{chan_corporate_2023}. In particular, physical closeness may enable tax offices to exert greater enforcement pressure~-- such as through door-to-door visits~-- which could increase compliance. Our analysis explores how these competing mechanisms play out by testing whether the effectiveness of behavioral nudges varies with business proximity to tax offices and across regions.

The economic literature has traditionally modeled tax evasion as a decision balancing the costs of detection and penalties against the benefits of evasion \citep{AllinghamSandmo1972,yitzhaki_income_1974}. 
More recently, attention has shifted to the moral and psychological dimensions of compliance, including social norms and moral costs \citep{luttmer_tax_2014-1}. Empirical evidence shows that deterrence interventions, such as audit notifications, tend to increase compliance \citep{slemrod_taxpayer_2001,mendoza_backfiring_2017}, while non-deterrence interventions~-- such as moral suasion or social norms messages~-- produce more mixed results \citep{blumenthal_normative_2001,torgler_importance_2006,hallsworth_behavioralist_2017}. Compliance is also shaped by fairness perceptions \citep{Tyler2006,Saundersetal2013,Besleyetal2019}, trust in authorities \citep{kirchler_enforced_2008}, and cognitive biases such as anchoring \citep{Maciejovskyetal2007}, framing \citep{AshbyWebley2008}, inertia \citep{Kerr2012,Jones2012}, and imperfect memory \citep{Ericson2017,gillitzer_nudging_2020}.

A growing body of work examines behavioral tax compliance in upper-middle-income countries, mainly in Latin America.\footnote{See, among others, \citet{eguino_nudging_2020}, \citet{schachtele_improving_2022}, \citet{schachtele_fiscal_2023}, \citet{ortega_deterrence_2013}, \citet{mogollon_whos_2021}, \citet{kettle_behavioral_2016}, \citet{kettle_failure_2017}, \citet{castro_tax_2013}, \citet{brockmeyer_casting_2019}, and \citet{holz_100_2023}.} 
In Indonesia, \cite{persian_behavioural_2023} show that email prompts modestly increase filing rates in a Randomized Controlled Trial (RCT) with more than~11 million taxpayers. The closest study to ours, a \cite{worldbank2020_rct} RCT with~24,000 MSMEs, finds that calendar and informational messages raised tax payments by~12\%, but combining these with deterrence messages increased filing rates only slightly and had no effect on payment amounts. 

We contribute to the behavioral tax compliance literature with a particular focus on developing country contexts. First, we test the effectiveness of three low-cost, behaviorally informed treatments~-- letters and flyers~-- to assess which most effectively improves compliance. Cost-effectiveness is central to scaling interventions, and our treatments are roughly two to three times cheaper than those studied in the \cite{worldbank2020_rct} report. Second, we examine how business location conditions responsiveness to behavioral interventions, an issue that has received limited attention. While earlier studies link proximity to local tax offices with compliance, ours is the first to investigate this relationship in an experimental setting. Third, we broaden the geographic scope by comparing metropolitan and non-metropolitan regions across Indonesia (Jakarta, South Sumatera, and West Nusa Tenggara), providing new evidence on how regional factors shape the effectiveness of behavioral nudges. Finally, we estimate heterogeneous treatment effects by penalty and filing history, showing how past behavior mediates responsiveness to nudges and informing the design of more targeted interventions.

Our results show that, on average across regions, all three treatments significantly improve compliance, raising filing rates, non-zero returns, and year-on-year growth in both payment likelihood and amount. Among the interventions, the deterrence letter has the strongest effect~-- raising non-zero filing by nearly~40 percentage points and increasing monthly tax payments by around~IDR40,000 (\$US2.75) per business. The information and tax revenue letters (including flyers) also yield positive, albeit more modest, impacts. 

More nuanced patterns emerge when treatment effects are disaggregated by region. In South Sumatera and West Nusa Tenggara~-- areas with lower baseline compliance~-- all three letters significantly increase both filing and payment rates, with the deterrence letter again proving most effective. In South Sumatera, this intervention brings compliance levels close to those observed in Jakarta. By contrast, in Jakarta~-- where compliance is already high, particularly among businesses located close to tax offices~-- our interventions reduce compliance, especially within~500 metres of a tax office. These negative effects weaken with distance and eventually turn positive. These findings highlight the importance of tailoring tax interventions to local compliance conditions. While simple deterrence messages can be highly effective in lower-compliance areas, they may be counterproductive in already compliant urban environments. 

The remainder of this paper is structured as follows. Section~2 provides an overview of the related literature and discusses potential mechanisms and expected behaviors. Section~3 presents the experimental design, provides a description of the data, and explains our estimation strategy. The results are presented in Section~4. Section~5 concludes.

\section{Related literature, mechanisms and expected behaviors}

\subsection{Related literature}

\noindent The tax compliance literature has expanded rapidly with the availability of administrative data and growing interest in testing cost-effective interventions to improve compliance \citep{hallsworth_behavioralist_2017}. Tax compliance decisions are shaped by both economic and non-economic considerations. The traditional framework models evasion as a choice comparing the expected benefits of under-reporting with the probability of detection and associated penalties \citep{AllinghamSandmo1972,yitzhaki_income_1974}. Consistent with this framework, deterrence interventions such as audit notifications are generally effective. For example, \cite{Slemrodetal2001} show that informing low- and middle-income earners about audit selection increases reported income, while high-income earners with greater evasion opportunities reduce their reported income. \cite{mendoza_backfiring_2017} document that compliance improves only up to a certain audit intensity, after which it declines.

A second key strand emphasizes the moral and social costs of non-compliance \citep{ErardFeinstein1994,Reckersetal1994,BobekHatfield2003,Torgler2007,AlmTorgler2011}, which arise when individuals deviate from social norms \citep{Elster1989,MylesNaylor1996,Wenzel2004,FreyTorgler2007,Traxler2010,Bobeketal2013}. However, evidence on non-deterrence interventions is mixed. Many studies find limited effects of moral suasion and social norm messages
\citep{blumenthal_normative_2001,Torgler2004,Wenzel2005,Fellneretal2013}, and a meta-analysis concludes tax morale messages are generally ineffective compared to neutral messages \citet{Antinyan2020}. Some interventions can even backfire, such as rewards that reduce compliance \cite{Dwengeretal2016}. Nonetheless, a few large-scale field experiments show positive effects of social norm messages on tax revenue \citep{hallsworth_behavioralist_2017,Bottetal2020}.

The costs and benefits of tax compliance can be influenced by various other factors. Taxpayers derive utility from public goods, affecting compliance through reciprocity \citep{BazartBonein2014} and altruism \citep{Andreoni1990,FeldFrey2010}. Compliance is further shaped by perceptions of fairness \citep{Tyler2006,Saundersetal2013,Besleyetal2019}, trust in authorities \citep{kirchler_enforced_2008}, interactions with tax administrations \citep{Braithwaite2009}, and behavioral biases such as anchoring \citep{Maciejovskyetal2007}, framing \citep{AshbyWebley2008}, inertia \citep{Kerr2012,Jones2012}, and imperfect memory \citep{Ericson2017,gillitzer_nudging_2020}.\footnote{\cite{BiddleArcos-Holzinger2016} provide a more detailed review of the determinants of tax compliance.}

Several studies have examined tax compliance in upper-middle-income countries such as Indonesia.\footnote{Only a few studies have explored the effectiveness of deterrence and non-deterrence nudges in lower-middle- and low-income countries. Among lower-middle-income countries, research has been conducted in Eswatini \citep{santoro_encouraging_2024}, Jordan \citep{alasfour_determinants_2016}, Pakistan \citep{slemrod_tax_2019}, Papua New Guinea \citep{hoy_improving_2021}, Tanzania \citep{collin_property_2022}, Vietnam \citep{nga_tax_2023} and Zimbabwe  \citep{dlamini_determinants_2017}. Important examples of studies in low-income countries include those by \cite{Mascagni2018} and \cite{mascagni_tax_2021} in Rwanda.} The majority of these studies focus on Latin American countries. In Argentina, \cite{eguino_nudging_2020} show that redesigned tax bills appealing to reciprocity raise payment rates by about~20\%~-- and nearly~40\% when delivered in person~-- with effects attributed to improved design, reciprocity, and proximity to public services. \cite{schachtele_fiscal_2023} find that reciprocity appeals have lasting impacts, with effects persisting two years later. Direct interactions with taxpayers also prove effective in Colombia: \cite{OrtegaSanguinetti2013} show that visits outperform letters and emails, while \cite{mogollon_whos_2021} document that phone calls increase tax collections by~25 percentage points.

RCTs in Guatemala further highlight the mixed effects of communication strategies. \citet{kettle_behavioral_2016} find that deterrence and social norm letters triple tax receipts in an experiment with~40,000 taxpayers, whereas other messages only increase declaration rates. By contrast, an online experiment with~600,000 taxpayers found no effect of honesty declarations, public goods messages, or penalty reminders on reported amounts \citep{kettle_failure_2017}. Evidence on property tax compliance also shows heterogeneity: deterrence messages raise compliance in Argentina \citep{castro_tax_2013}, while fairness appeals are ineffective; registration nudges in Brazil yield positive effects but lottery rewards backfire \citep{schachtele_improving_2022}; social norm messages in Peru boost compliance by 20\% \citep{DelCarpio2022}; and enforcement in Mexico raises revenue but reduces welfare \citep{brockmeyer_taxing_2021}.

Evidence on business tax compliance in upper-middle-income countries is more limited. In Costa Rica, \citet{brockmeyer_casting_2019} find that enforcement emails raise tax payments by~3.4 percentage points among non-filing firms, with third-party reporting further improving compliance. In the Dominican Republic, deterrence messages stressing prison sentences and public disclosure, combined with enforcement reforms, raised tax revenues by~US\$184 million, with especially strong effects among large firms \citep{holz_100_2023}.

We are aware of only two experimental studies examining behavioral aspects of tax compliance in Indonesia. \citet{persian_behavioural_2023} evaluate email prompts in an RCT with more than~11 million taxpayers and show that all treatments raise early and overall filing, with planning prompts most effective. The closest study to ours is a \citet{worldbank2020_rct} RCT involving~24,000 MSMEs in Java (excluding Jakarta), which tested calendars paired with informational, public goods, or deterrence messages. Payment effects were modest: information and public goods treatments raised payments by about IDR13,000 (US\$0.89),\footnote{Throughout the paper, we use an exchange rate of US\$1 = IDR14,528 as of 1~July~2021 (\url{https://www.xe.com/currencytables/?from=USD\&date=2021-07-01\#table-section}, accessed on 15 January 2025). This rate is comparable to the exchange rates during the study period of the \cite{worldbank2020_rct} report.} while deterrence increased payment rates by one percentage point without affecting amounts. 

A simple cost-benefit analysis reveals that our treatments are more cost-effective than those used by the World Bank. For every dollar spent, the World Bank's most effective treatment generated about~US\$3.41 in tax revenue over a~12-month period and about~US\$4.22 over the~15-month period observed in the data.\footnote{The effects presented in Table 4.3, Column~4 of the \cite{worldbank2020_rct} report are: Information: IDR12,879 (US\$0.89), Public goods: IDR13,004 (US\$0.90), Deterrence: IDR$-$2,945 (US\$$-$0.20, not statistically significant). The cost of printing and shipping a calendar, as shown in Table~4.7 of the report, is IDR44,197 (US\$3.04). Using the discount rate of~6.66\% per annum employed in the report, the present value of monthly tax payments of IDR13,004 (US\$0.90) over a 12-month period is IDR150,561 (US\$10.36). The resulting cost-benefit ratio is 44,197/150,561=3.04/10.36=0.294. In other words, for every dollar spent, the amount of tax revenue collected is~US\$3.41. The corresponding amount of tax collected over a 15-month period is IDR186,665 (US\$12.85), yielding a tax revenue of US\$4.22 for every dollar spent.} In contrast, our treatments were considerably cheaper and, in the case of the deterrence letter, far more effective. For every dollar spent, our deterrence letter generated about~US\$30 in tax revenue over a~12-month period. We present the results of our cost-benefit analysis in Section~4.3 below.

\subsection{Mechanisms and expected behaviors}

\noindent To better understand the mechanisms underlying business tax compliance, we compare the costs and benefits of tax evasion. Specifically, we consider the deterrence model of \cite{AllinghamSandmo1972}, which extends \cites{becker_crime_1968} economic model of crime to the domain of tax evasion. Using the notation introduced by \cite{Slemrod2019}, we may write
\begin{eqnarray} \label{eq1} (1-p(e,a)) U(y(1-t) + te) + p(e,a) U(y(1-t) - fte), \end{eqnarray}
where the probability of detection, $p(e,a),$ is a function of the understated tax liability~($e$) and the enforcement intensity~($a$). The function $U(\cdot)$ is a von Neumann-Morgenstern (VNM) utility function. The variable~$y$ denotes the taxpayer's true income, $t$ is a proportional income tax rate, and $f$ is the penalty factor applied to detected evasion.

Applying the VNM theorem, equation~(\ref{eq1}) may be written as
\begin{eqnarray} \label{eq2} \mathbb{E} \big [U((1-p(e,a)) (y(1-t) + te) + p(e,a) (y(1-t) - fte)) \big ]. \end{eqnarray}

It is useful to consider the case of full compliance (no evasion). In this case, the expected utility reduces to
\begin{eqnarray} \label{eq3} \mathbb{E} [U( y(1-t))]. \end{eqnarray}

Comparing~(\ref{eq2}) and~(\ref{eq3}), taxpayers will be fully compliant if 
\begin{eqnarray} \nonumber  \underbrace{(1-p(e,a))te}_{\textnormal{expected benefit of evasion}}  &<& \underbrace{p(e,a) fte.}_{\textnormal{expected cost of evasion}}  \end{eqnarray}

Proximity to the local tax office affects our model through the probability of detection, $p(e,a)$. Several studies highlight that compliance is more likely influenced by the believed probability of detection, $p_b(e,a)$, rather than the actual probability, $p(e,a)$.\footnote{Examples include \cite{alm_estimating_1992}, \cite{snow_audit_2007}, \cite{dhami_why_2007} and \cite{durlauf_deterrent_2010}.} This distinction underscores the importance of taxpayer perceptions, which may diverge from objective realities.

If businesses located near a tax office perceive their proximity as providing an informational advantage~-- as suggested by \cite{kubick_irs_2017}~-- they may believe their probability of detection is lower than that of businesses located farther away: $p_b^{\textnormal{close}}(e,a) < p_b^{\textnormal{far}}(e,a)$. As a consequence, businesses closer to a local tax office would face a higher expected benefit from tax evasion and a lower expected cost of evasion compared to those located farther away. Conversely, if local tax offices are perceived as more effective at gathering information about nearby businesses than those businesses are at obtaining information about local tax offices~-- as suggested by  \citep{chan_corporate_2023}~-- then the believed probability of detection for nearby businesses would exceed that of businesses farther away: $p_b^{\textnormal{close}}(e,a) > p_b^{\textnormal{far}}(e,a)$. In this scenario, the outcomes are reversed: businesses closer to a local tax office would face a lower expected benefit from tax evasion and a higher expected cost compared to their more distant counterparts. 

While the theoretical model suggests that enforcement messages may enhance compliance by raising the believed probability of detection, it remains unclear whether these messages will contribute to closing or widening the compliance gap between businesses located near local tax offices and those farther away. This insight underscores the importance of empirical analysis in determining the link between business proximity to local tax offices and their responsiveness to deterrence messaging. 

In addition to using enforcement messages, we also examine the effectiveness of providing information aimed at facilitating business tax compliance. From a theoretical perspective, providing information to reduce compliance costs should affect all businesses, except those that do not comply at all. We extend equation~(\ref{eq2}) to account for compliance costs,~$\delta,$ which are assumed to increase with reported income,~$r=y-e$: 
\begin{eqnarray} \label{eq4} \mathbb{E} \big [U((1-p(e,a)) (y(1-t) + te - \delta r ) + p(e,a) (y(1-t) - fte  - \delta r )) \big ]. \end{eqnarray}

By comparing equation~(\ref{eq4}) to the case of full compliance, we obtain 
\begin{eqnarray} \nonumber  \underbrace{(1-p(e,a))te + \delta e}_{\textnormal{expected benefit of evasion}}  &<& \underbrace{p(e,a) fte.}_{\textnormal{expected cost of evasion}}  \end{eqnarray}

This result suggests that when compliance costs rise with reported income, they increase the expected benefit of evasion. Our information treatment aims to enhance compliance by reducing compliance costs. However, the effectiveness of providing information may vary depending on its usefulness to businesses, a factor not addressed in the model. If businesses in metropolitan areas~-- typically located near local tax offices~-- are better informed than those in non-metropolitan areas, we would expect the provision of information to be particularly useful for businesses in non-metropolitan areas. At the same time, businesses in rural areas may benefit less from information about digital solutions that require internet access, such as guidance on using apps or accessing online resources, due to limited internet connectivity.

Finally, we assess the effectiveness of a treatment designed to increase awareness of how tax revenue is used. Providing information on the use of tax revenue has the potential to enhance tax compliance because taxpayers may benefit from the provision of public goods \citep{cowell_unwillingness_1988}, with reciprocity \citep{BazartBonein2014} and altruism \citep{Andreoni1990,FeldFrey2010} serving as potential underlying mechanisms. For simplicity, we assume that the benefits derived from the provision of public goods, captured by the parameter~$\theta$, increase with reported income, which determines the amount of tax paid. We incorporate the utility gain from public goods into our model by extending equation~(\ref{eq4}) as follows:
\begin{eqnarray} \label{eq5} \mathbb{E} \big [U((1-p(e,a)) (y(1-t) + te - \delta r + \theta r ) + p(e,a) (y(1-t) - fte - \delta r + \theta r )) \big ]. \end{eqnarray}

Comparing equation~(\ref{eq5}) to the case of full compliance yields 
\begin{eqnarray} \nonumber  \underbrace{(1-p(e,a))te + \delta e}_{\textnormal{expected benefit of evasion}}  &<& \underbrace{p(e,a) fte + \theta e.}_{\textnormal{expected cost of evasion}}  \end{eqnarray}

This result suggests that providing information about the use of tax revenue increases the expected cost of tax evasion because taxpayers may derive additional utility from compliance when they are aware of the benefits of public goods and their role in funding them. The extent to which the effects of providing information about the use of tax revenue vary across business locations remains an empirical question. Our trial was conducted during the Covid-19 pandemic in 2021 and highlighted the use of tax revenue to fund vaccinations~-- a public good with immediate and widespread relevance. It is plausible that this type of information is particularly effective in areas where public health measures have visible impacts or where trust in government initiatives is higher.

\section{Trial design, data and empirical strategy}

\subsection{Trial design}

\noindent Our trial examines MSME tax filing and payment behavior across three regions: Jakarta, South Sumatera, and West Nusa Tenggara.\footnote{We use the terms ``trial'' and ``natural field experiment'' interchangeably throughout this paper. Natural field experiments are typically defined as RCTs in which participants are unaware that they are part of the study \citep{Cziboretal2019}.} The trial was designed in collaboration with the DGT and implemented through~13 local tax offices. The sample was drawn directly from these offices, which determined the geographic scope of the study. The sampled business locations are shown in Appendix~A. In Jakarta, all participating businesses are located in North Jakarta, a densely populated metropolitan district. In South Sumatera, the sample covers both the city of Palembang and the regional town of Prabumulih, including a mix of urban, semi-urban, and rural areas. Similarly, in West Nusa Tenggara, the sample includes businesses in and around the city of Mataram and nearby regional towns, again capturing a diverse mix of settlement types.

MSMEs in Indonesia are required to file tax returns and make tax payments on a monthly basis.\footnote{Article~7(3) of the 2023 Ministry of Finance Regulation. Payment receipts may serve as tax returns for payments validated with a state revenue transaction number (Nomor Transaksi Penerimaan Negara, or NTPN).} Failure to submit tax returns or make timely payments may result in an administrative penalty of IDR100,000~(US\$6.88), in addition to a~2 percent interest charge on the outstanding amount. The current tax rate for MSMEs is~0.5 percent, applicable to businesses with an annual turnover of up to IDR4.8 billion (US\$330,396), in line with the definition of MSMEs under Indonesian tax regulations.\footnote{The~0.5 percent rate under GR-23/2018 replaced the previous~1.0 percent rate from GR-46/2013. The reduction was implemented to provide greater financial relief to MSMEs, enabling them to retain more income for business expansion. The new rate took effect on 1~July~2018.}

Our analysis uses administrative records from the DGT, compiled monthly from anonymized information submitted by local tax offices. All data were de-identified in accordance with Article~34 of Indonesia's General Provision and Tax Procedure Law, which mandates strict confidentiality and anonymity. Ethics clearance was obtained from the Human Research Ethics Committee of the Australian National University.\footnote{ANU Human Research Ethics protocol number 2021/508.} The target population comprised approximately~12,000 registered MSMEs~-- defined as businesses with an annual turnover of less than IDR4.8 billion (US\$330,396)~-- recorded by the tax office between~2017 to~2019.\footnote{The time frame was chosen to capture the application of prior regulations, such as GR-46/2013, while avoiding selection bias related to special arrangements due to COVID-19 in 2020.} Additionally, we focus on MSMEs classified by the DGT as ``monitored'' taxpayers~-- those who submitted zero tax returns or filed late on at least three occasions within the~12 months prior to the intervention. Registered MSMEs in Indonesia represent a highly selective subgroup of small businesses. As of 2022, there were about~2.3 million registered small businesses \citep{dgt_smes_2022}, compared to an estimated~63 million MSMEs in total \citep{indonesia_ministry_of_smes}, meaning that only about~3.7 percent of MSMEs are officially registered.

Our trial compares members of three treatment groups to members of a control group who did not receive a letter. The treatment letters (presented in Appendix~B) were designed to target three distinct behavioral aspects of tax compliance. The first letter, the ``Deterrence Letter'', sought to reinforce compliance through clear reminders of submission deadlines, the administrative penalty of~IDR100,000 (US\$6.88) for late filing, and the increasing use of computerized audits. It also emphasized the potential for further enforcement actions against non-compliance. The second, the ``Literacy Letter'', adopted a simplified and approachable tone, providing step-by-step instructions to help recipients fulfill their tax obligations. It included a QR code that directed recipients to the official mobile application \textit{m-pajak} (see Appendix~B for details) and the DGT website, offering easy access to additional resources. The third, the ``Public Goods Letter'', aimed to inspire compliance by emphasizing the positive societal impacts of tax payments, such as funding COVID-19 relief efforts and supporting education for future generations. To enhance the effectiveness of the Literacy and Public Goods letters, we included informative flyers designed with color and graphic elements grounded in psychological research on visual perception and engagement.\footnote{The use of color psychology in treatment design was informed by \cite{elliot_color_2015}. Previous studies have shown that red stimuli attract attention and convey caution \citep{lindsay_meeting_2010, tchernikov_color_2010, buechner_red_2014, Pomerlau2013, Sokolik2014}, while blue and yellow light have been linked to increased alertness and improved performance on attention-based tasks \citep{lockley_short-wavelength_2006, viola_blue-enriched_2008, cajochen_evening_2011, taillard_-car_2012}.}

Before the trial, the DGT provided pre-treatment characteristics for~11,996 taxpayers. To ensure stratified randomization, we grouped taxpayers with similar pre-treatment characteristics into strata. Within each stratum, cases were randomly assigned to one of the treatment groups or the control group. This approach ensured an even distribution of taxpayers across groups while maintaining balance in pre-treatment characteristics. The randomization process was carried out using a random number generator in \texttt{STATA\textsuperscript{\textregistered}}, with a randomly selected seed to ensure replicability. Specifically, we employed the user-written command \texttt{randtreat} (version 1.4) for stratified randomization, which is designed to facilitate random assignment within predefined strata. The random assignment resulted in the following allocation of taxpayers: Treatment Group~1 (Deterrence) with 2,991 taxpayers; Treatment Group~2 (Information) with 2,989 taxpayers; Treatment Group~3 (Public Goods) with 3,040 taxpayers; and the Control Group with 2,976 taxpayers.

Our trial was implemented in November 2021, with letters sent out throughout the month. Outcomes were measured monthly from January 2022 to February 2023 (a total of~14 months), with data collection concluding in February 2023.\footnote{We restrict our analysis sample to the first~12 months after the intervention because we only received information about changes in tax payments relative to the period~12 months earlier. We discuss this issue in greater detail in Section~3.2 below.} The trial was registered with the Social Science Registry of the American Economic Association on 18~May~2022 (RCT ID AEARCTR-0009460). The pre-analysis plan was submitted on 17~February~2023, and we received access to the post-trial data in March 2023. While our pre-analysis plan focused on tracking the evolution of outcomes over time, we decided to exclude dynamic aspects from our analysis due to limited variation in outcomes over time. Instead, we pooled outcomes across the period and clustered standard errors to account for repeated observations over time. We also streamlined the definitions of outcome variables and considered only two of the five dimensions covered in the pre-analysis plan: filing and payment. Specifically, we focus on four primary outcomes: whether a business submitted a tax return, whether the return was non-zero, whether there was an increase in payment compared to the previous year, and the amount by which payments increased year-on-year. Other outcomes listed in the pre-analysis plan~-- such as timely submission, use of electronic payments, and inbound calls to the tax office~-- were not included in our analysis, as they were of secondary importance.

A common concern in the design of field experiments is the potential for spillover effects arising from interactions between members of the treatment and control groups. Specifically, contamination of the control group could bias the estimated treatment effects, which are based on comparisons between treatment and control groups. While we cannot entirely rule out the possibility of spillovers, it appears unlikely that they represent a significant issue in our study. Our analysis sample consists of MSMEs spread across three distinct islands, with a wide geographic spread and a combined population of over~200 million.

\subsection{Data}

\noindent The data provided by the DGT prior to the trial implementation were used for stratified randomization. Following the interventions, external data sources were incorporated to add two baseline variables to the DGT's administrative records: one measuring the distance to the nearest tax office\footnote{Calculated as the Euclidean distance between the taxpayer's registered location and the nearest tax office, based on their longitude and latitude coordinates, the measure represents the straight-line distance (as the crow flies) between two points on a Cartesian plane.} and another capturing the base transceiver station (BTS) density.\footnote{BTS density refers to the number of mobile network towers in a given area (regency or city), which can influence internet accessibility and connectivity for businesses. Unfortunately, the 2007 measure from the Ministry of Communications and Information, which was available at the time of our trial implementation, omits data from more recent cellular operators. Validation with an updated BTS density measure from Geospatial Information Indonesia Statistics (2020) confirmed that the 2007 measure is outdated and unreliable. Therefore, we do not consider this variable in our analysis of heterogeneous treatment effects.} The set of pre-treatment characteristics used for stratified randomization includes three regions (Jakarta, South Sumatera, and West Nusa Tenggara), two sectors (service, trade/industry), two business types (business-only taxpayers, business taxpayers with additional income), two tax return groups (individuals, firms), and sextiles of business turnover, business age, distance to the nearest tax office, and base transceiver station (BTS) density.

The DGT provided access to further pre-treatment characteristics at the end of the data collection period after the trial. Specifically, we obtained information on whether businesses had received penalties and the number of months they had failed to file in the~12 months prior to the treatment. We used the latter to create a variable indicating whether businesses had failed to file for more than six months, which may be viewed as a risk indicator from the tax authority's perspective. These additional variables allow us to study heterogeneous treatment effects based on penalty and filing history. 

Summary statistics of the pre-treatment variables are presented in Table~1. The $p$-values in Table~1 correspond to comparisons of sample means between each treatment group and the control group. Most of these comparisons show no significant differences, confirming that the randomization process effectively balanced the covariates across groups. However, we observe a significant difference ($p$-value $=$ 0.024) between Treatment Group~3 and the Control Group with respect to penalty history, which was not used for stratified randomization. To account for imbalances, we will include the pre-treatment variables presented in Table~1 as control variables in our regression analysis.

\begin{center}
	[Table 1 about here.]
\end{center}

We use the administrative records provided by the DGT at the end of the data collection period to construct four outcome measures: (1) a binary variable indicating whether a tax return was filed, (2) a binary variable indicating whether a non-zero tax return was filed, (3) a binary variable indicating a higher amount of tax paid relative to the previous year, and (4) the additional amount of tax paid (in IDR) relative to the previous year. 

The DGT's compliance risk management system focuses on tracking increases in tax payments compared to the same period~12 months earlier, as most registered MSMEs either increase their tax payments or keep them stable. As a result, we do not observe tax payments in absolute terms, either before or after the intervention. This limitation has two implications. First, due to the comparison of tax payments relative to tax payments made~12 months earlier, analyzing the entire 14-month period following the intervention may create inconsistencies. We address this problem by restricting our analysis to the first~12 months after the intervention. Second, while we received data indicating whether MSME tax payments declined compared to the previous year, we have no information on the extent of that decline. However, only~56 of the~11,996 MSMEs in our dataset (0.047 percent) reduced their tax payments. Of those, 16 MSMEs belong to Treatment Group~1, 15 to Treatment Group~2, 9 to Treatment Group~3, and~16 to the Control Group. We exclude these~56 MSMEs from our analysis sample. Given the small number of cases and their fairly even distribution across groups, it appears plausible that their exclusion has a negligible impact on our results.

\subsection{Empirical strategy}

\noindent Our analysis assesses the impact of the three treatments on the four outcome measures introduced in the previous section. We pool monthly observations over the~12 months following the intervention and compute clustered standard errors to account for repeated observations over time. Our primary estimation strategy relies on linear regression with covariate adjustment. Specifically, we estimate the equation
 \begin{eqnarray} \label{eq6}
Y_{it} &=& \alpha_{0} + \beta_1 T^{1}_{i} + \beta_2 T^{2}_{i} +\beta_3 T^{3}_{i} +X_{i} \Gamma + \delta_{t}+\varepsilon_{it},
\end{eqnarray}
where~$Y_{it}$ denotes the outcome for MSME~$i$ ($i=1,\ldots, N$) in month $t$ ($t=1, \dots, 12$) post-intervention. $T^{1}_{i}$ $T^{2}_{i}$, and $T^{3}_{i}$ are binary indicators for assignment to the respective treatment groups, while $X_{i}$ is a vector of baseline covariates. $\delta_{t}$ captures month fixed effects, and~$\varepsilon_{it}$ is the error term. Our parameters of interest are $\beta_1$, $\beta_2,$ and $\beta_3$, which represent the average treatment effects on the treated.\footnote{These parameters can also be interpreted as average intent-to-treat (ITT) effects, as we cannot verify whether business owners received or read the letters. Although the possibility of returned mail cannot be entirely ruled out, delivery was made to the taxpayer's registered address as listed in the DGT database. According to the DGT, this database is updated quarterly, resulting in a low rate of undeliverable mail~-- fewer than 10\% of letters are returned to sender.} We use our regression results to conduct pairwise tests to compare treatment effects across groups. 

To explore heterogeneous treatment effects, we estimate interacted models that allow treatment effects to vary across predefined subgroups (e.g., by region). The model specification is as follows:
\begin{eqnarray} \label{eq7}
Y_{it} &=& \alpha_{0} + \sum_{j \in \mathcal{J}}\beta_1^{j} T^{1}_{i} G^{j}_{i} +\sum_{j \in \mathcal{J}}\beta_2^{j} T^{2}_{i} G^{j}_{i}+\sum_{j \in \mathcal{J}}\beta_3^{j} T^{3}_{i} G^{j}_{i}+X_{i} \Gamma + \theta_j + \delta_{t}+\varepsilon_{it},
\end{eqnarray}
where $G^{j}_{i}$ is a binary indicator equal to~1 if individual~$i$ belongs to group $j$, and~0 otherwise. The $\theta_j$ are group fixed effects. Our key parameters of interest in this specification are the $\beta_1^{j}$s, $\beta_2^{j}$s, and $\beta_3^{j}$s, which capture treatment effects within each subgroup.\footnote{When analyzing heterogeneity by baseline characteristics within regions, we estimate equation~(\ref{eq7}) separately for each region.}

\section{Results}

\subsection{Treatment effects}

\noindent Table~2 presents the treatment effects across three groups~-- T1 (Enforcement), T2 (Information), and T3 (Use of tax revenue)~-- compared to the control group, with results for both filing and payment outcomes. For filing behavior, all treatment groups show significant positive effects on tax filing compared to the control group. Specifically, T1 and T2 have the largest effects, increasing tax filing rates by~23 to~24 percentage points relative to the control group. T3 also shows a positive effect, though smaller in magnitude, with an increase of about~9 percentage points. Regarding non-zero filing, T1 has the largest effect, boosting non-zero filing by about~40 percentage points. In contrast, T2 and T3 show much smaller effects, ranging from about~8 to~12 percentage points. The effect of T1 on non-zero filing is significantly larger than those of T2 and T3, highlighting that T1 is particularly effective in improving MSME tax compliance.

\begin{center}
	[Table 2 about here.]
\end{center}

Turning to payment outcomes, all treatment groups show significant positive effects on the likelihood of increasing tax payments compared to the previous year. T1 again shows the largest effect, increasing the likelihood of higher payments by nearly~43 percentage points, followed by T3 at about~13 percentage points, and T2 at~9 percentage points. Similar to the non-zero filing results, the effect of T1 on year-on-year payment increases is significantly larger than those of T2 and T3, confirming that enforcement is the most effective tool for collecting tax revenue. Regarding increases in tax payments, T1 again shows the largest effect, causing an increase of~IDR39,379 (US\$2.71) per month, followed by T2 at IDR5,565 (US\$0.38) and T3 at IDR3,812 (US\$0.26). Taken together, these findings reveal that T1 has a substantial and consistent positive impact on both filing and payment, making it the most effective intervention among the three.

\subsection{Effects heterogeneity}

\noindent Our analysis of heterogeneous treatment effects examines how business location influences tax compliance, with a particular focus on the geographic proximity of businesses to local tax offices. We begin by exploring heterogeneous treatment effects by region in Section~4.2.1 before turning to the effects of distance to the local tax office within each region in Section~4.2.2. In Section~4.2.3, we summarize the main findings from our analysis of heterogeneous treatment effects across several other subgroups.

\subsubsection{Effects by region}

\noindent Table~3 presents the treatment effects by region. In Jakarta, all treatments have significant negative effects on the likelihood of tax filing, with T1 (Enforcement) showing the largest negative effect of about~10 percentage points compared to the control group. On the other hand, in South Sumatra and West Nusa Tenggara, all treatments show significant positive effects, with T1 having the largest effect in South Sumatra, increasing tax filing by about~77 percentage points, and in West Nusa Tenggara, increasing tax filing by about~48 percentage points. Regarding non-zero filing, in Jakarta, T1 shows the largest positive effect (about~31 percentage points), followed by T3 (Use of tax revenue) at about~12 percentage points, while the effect of T2 (Information) is statistically insignificant. In South Sumatera, T1 again shows a substantial positive effect (about~62 percentage point), while T2 and T3 yield more moderate positive effects (16-22 percentage points). In West Nusa Tenggara, T1 shows a positive effect of about~27 percentage points, while T2 and T3 show smaller effects (4-8 percentage points). 

\begin{center}
	[Table 3 about here.]
\end{center}

For the outcome measuring whether businesses increased their tax payments compared to the previous year, T1 consistently shows significantly positive effects across all regions. The largest effect is observed in South Sumatera (approximately~64 percentage points), followed by Jakarta (around~34 percentage points) and West Nusa Tenggara (about~32 percentage points). T2 and T3 also produce positive and statistically significant effects in South Sumatera~-- ranging from~17 to~23 percentage points~-- and in West Nusa Tenggara, where effects range from~5 to~9 percentage points. In Jakarta, T3 shows a moderate effect of approximately~12 percentage points, while the effect of T2 is not significant. 

Turning to the actual increase in tax payments, T1 again generates the largest gains, particularly in South Sumatera, where the monthly increase in payments amounts to IDR55,421 (US\$3.82). Smaller but still meaningful increases are observed in Jakarta (IDR33,874, or US\$2.33) and West Nusa Tenggara (IDR25,090, or US\$1.73). While T2 and T3 also lead to increased payments in South Sumatera, the magnitudes are considerably smaller than those of T1. In Jakarta and West Nusa Tenggara, the effects of T2 and T3 on payment amounts are positive but more modest and not always significant.

The underlying drivers of these regional differences in treatment effects remain uncertain. The three regions vary in terms of economic development, population density, and a range of geographic, cultural, and administrative characteristics, all of which could plausibly influence compliance behavior and responsiveness to interventions. While we do not aim to provide a comprehensive explanation of the regional variation, our heterogeneity analysis highlights one key factor: distance to the local tax office. This variable not only differs substantially across regions but is also strongly correlated with both baseline compliance and treatment effects.

\subsubsection{Effects by distance to the local tax office}

\noindent We find substantial variation in the distance to local tax offices across the three study regions. As shown in Table~4, in Jakarta (specifically North Jakarta district), 50\% of businesses in our sample are located within one kilometer of the nearest local tax office. In contrast, in South Sumatera, 90\% of businesses are more than six kilometers away, and in West Nusa Tenggara, 90\% are located more than 23 kilometers from the nearest tax office. These differences reflect both disparities in population density and the limited presence of tax offices in more remote areas. 

\begin{center}
	[Table 4 about here.]
\end{center}

When we examine compliance behavior in the control group by distance to the local tax office, disaggregated by region, distinct patterns emerge. In areas extremely close to a tax office~-- such as the first decile of the distance measure in Jakarta, where businesses are located within~500 meters~-- compliance is nearly perfect. Filing rates in this group reach~100\%, and average monthly tax payments increased by IDR64,000 (US\$4.41) compared to the previous year.
This increase exceeds inflation and is substantial given the average annual turnover and tax liability of MSMEs in our Jakarta sample. These businesses are not only consistently filing and reporting non-zero returns but are also showing a marked increase in their payments over time.\footnote{It is important to keep in mind that registered MSMEs represent a highly selected subset of MSMEs in Indonesia. As such, compliance behavior among this group is not representative of MSMEs more broadly. In a context like Indonesia, where verifying MSME turnover poses administrative challenges, the compliance behavior observed among registered MSMEs located within~500 meters of a tax office in Jakarta can be considered an upper bound of what is achievable in terms of MSME compliance.}

While we cannot conclusively determine the mechanisms behind this high level of compliance, administrative records from the DGT and prior research suggest that proximity to tax offices may facilitate more frequent site visits by tax officials~-- an enforcement and outreach strategy commonly used in Indonesia \citep{Antonacci2024, Basri2021, dgt_doorknocking_2015}. These visits may improve taxpayers' understanding of filing obligations or increase their perceived risk of detection and penalties for noncompliance. Interestingly, tax compliance in Jakarta declines as the distance from the nearest local tax office increases~-- particularly in terms of non-zero filing and increased payments. Among businesses located more than one kilometer away from the tax office, fewer than 10\% file non-zero returns, representing a nearly 90 percentage point drop compared to those within~500 meters. Similar patterns are observed in South Sumatera and West Nusa Tenggara. However, because the baseline compliance is lower in these regions, the decline with distance is less pronounced than in Jakarta.

Figures~1 and~2 present treatment effects by deciles of distance to the nearest tax office for each region. The first notable finding is that, in Jakarta, treatments backfire among businesses located very close to a tax office~-- reducing both filing rates and tax payments. This negative effect dissipates after the second or third decile, depending on the outcome. Beyond that point, treatment effects generally improve with distance, and in higher deciles, compliance levels among treated businesses approximate those of firms located near tax offices. T1 consistently produces the strongest effects, especially for increased payments. Figure~3 presents a map of treatment effects in North Jakarta, illustrating the presence of six local tax offices situated in close proximity within a relatively compact area. This dense clustering increases the likelihood of interactions between taxpayers and tax officials. 

\begin{center}
	[Figure 1 about here.]
\end{center}

The second key result is that, in South Sumatera, all treatments~-- particularly T1~-- are noticeably more effective. The large magnitude of the effect of T1 observed in South Sumatera suggests that T1 raises overall compliance to a level comparable to that of businesses within~500 meters of a tax office in Jakarta~-- our high-compliance benchmark. This is a striking result, especially given that the intervention consisted of a simple deterrence letter. In other words, in a setting with lower baseline compliance with fewer enforcement resources, a low-cost behavioral intervention is able to raise tax compliance to levels typically observed in a highly monitored, metropolitan area. This finding highlights the considerable potential of such interventions in improving compliance in under-resourced regions. 

\begin{center}
	[Figure 2 about here.]
\end{center}

The third insight relates to the interaction between distance and treatment heterogeneity across regions. While greater distance is associated with stronger treatment effects in Jakarta, no consistent pattern is observed in South Sumatera and West Nusa Tenggara. This suggests a non-linear relationship between distance and responsiveness to compliance interventions. Within a certain proximity~-- approximately 1.5 kilometers in our case~-- treatments may be more effective among businesses further away from the tax office, perhaps due to their initially lower compliance levels. However, at much greater distances (over~6 kilometers in South Sumatera and~22 kilometers in West Nusa Tenggara), the relationship appears to weaken. In these more remote areas, the limiting effect of remoteness itself~-- potentially due to infrastructure, access to information, or administrative capacity~-- may constrain the effectiveness of interventions. 

\begin{center}
	[Figure 3 about here.]
\end{center}

\subsubsection{Additional sources of heterogeneity}

\noindent We also estimate heterogeneous treatment effects across subgroups based on the remaining pre-treatment characteristics presented in Table~1, separately for each region. We find no evidence of heterogeneity by turnover in Jakarta or West Nusa Tenggara. However, in South Sumatera, the deterrence treatment has a significantly stronger effect on filing among businesses with higher turnover. Similarly, we observe no heterogeneity by business age in Jakarta and West Nusa Tenggara, but in South Sumatera, the deterrence letter leads to larger increases in filing, non-zero filing, and increased payments for businesses that have been operating for a longer period. 

When examining heterogeneity by past penalties and filing history, notable patterns emerge only in Jakarta. There, the deterrence letter has significantly stronger effects on filing, non-zero filing, and increased payments among businesses that had more than six months of non-filing in the previous year. The treatment effect of the deterrence letter is also larger for businesses with a history of penalties. These results suggest that in Jakarta, where baseline compliance is relatively high, the treatments are particularly effective among businesses that were previously less compliant. 

Our analysis of treatment effects across tax return groups, taxpayer types, and sectors reveals further heterogeneity, with patterns that are often mixed and challenging to interpret consistently across regions. In Jakarta, business-only taxpayers exhibit lower responsiveness to T1 (Enforcement) in both filing and payment behavior compared to business taxpayers with additional income sources. Moreover, businesses in the service sector are more responsive than those in the trade/industry sector. These patterns generally hold for T2 (Information) and T3 (Use of tax revenue), although the differences are smaller in magnitude and not always statistically significant.

In South Sumatera, we find that the effect of T1 on filing is slightly weaker in the service sector compared to the trade/industry sector. Furthermore, individual taxpayers are more likely than firms to increase their payments in response to T1. Other subgroup differences in this region are not statistically significant.

In West Nusa Tenggara, we observe more pronounced heterogeneity. Individual taxpayers are consistently more responsive than firms, business-only taxpayers respond more strongly than those with additional income, and the trade/industry sector is more responsive than the service sector. These differences are most evident for T1, less so for T2, and weakest~-- and not always statistically significant~-- for T3. One potential explanation for these findings is the relatively small sample size in West Nusa Tenggara compared to Jakarta and South Sumatera, which may contribute to greater variability and in the estimates and reduced statistical power.

\subsection{Cost-benefit analysis}

\noindent Table~5 presents the results of our cost-benefit analysis for the three treatments. Regarding the amount of tax paid, T1 (Enforcement) generates the highest monthly tax payment at IDR39,379 (US\$2.64), followed by T2 (Information) at IDR5,565 (US\$0.38) and T3 (Use of tax revenue) at IDR3,812 (US\$0.26). The present value of the annual amount of tax paid is IDR455,933 (US\$30.61) for T1, compared to IDR64,432 (US\$4.44) for T2 and IDR44,136 (US\$3.04) for T3.

Regarding costs, the printing cost per letter is IDR800 (US\$0.06) for both T1 and T3, and IDR1,000 (US\$0.07) for T2.\footnote{The printing cost per letter for T2 is higher due to the use of a non-standard, informal format. Unlike the standardized letters in T1 and T3, which follow a uniform design, T2 features a more visually engaging layout with additional colors to enhance readability and appeal. This customization requires the use of higher-quality printing materials.} In addition, both T2 and T3 incur extra costs of IDR3,000 (US\$0.21) for printing flyers. The postage cost per letter is consistent across all approaches, at IDR14,402 (US\$0.99). Consequently, the total costs amount to IDR15,202 (US\$1.05) for T1, IDR18,402 (US\$1.27) for T2, and IDR18,202 (US\$1.25) for T3.

The benefit-cost ratio is considerably higher for T1, at 29.99, indicating that for every dollar spent on T1, the present value of the amount of tax collected over a 12-month period amounts to US\$29.99. In contrast, for T2, each dollar spent yields US\$3.50, while for T3, the return is US\$2.42.

\section{Conclusions}

\noindent This paper presents findings from a randomized field experiment conducted in collaboration with the Indonesian tax authority, using administrative data to assess the effectiveness of behavioral interventions~-- specifically, hard-copy letters and flyers from local tax offices~-- in improving tax compliance among micro, small, and medium enterprises. Our sample consists of businesses already registered with the tax authority, and thus likely to exhibit a relatively high baseline compliance. However, we find substantial regional variation in baseline compliance: in metropolitan Jakarta, where businesses are typically located near multiple local tax offices, baseline compliance was significantly higher than in South Sumatera and West Nusa Tenggara, where businesses tend to be farther from tax offices. 

Treatment effects also vary substantially across regions. In Jakarta, where compliance is already high, especially among businesses located within close proximity to a local tax office, we find that the interventions backfired. In contrast, in South Sumatera, the deterrence letter proved highly effective, raising compliance levels to match those observed in high-compliance areas of Jakarta. These findings highlight the importance of geographic context in tax collection, suggesting that location-sensitive strategies may be essential for improving compliance in diverse settings. 

Our work contributes to a growing body of experimental research on tax compliance in low- and middle-income countries and represents one of the first field experiments targeting small businesses taxpayers in Indonesia. The results underscore the potential of simple, low-cost deterrence interventions in improving compliance, particularly in regions with lower baseline compliance and among businesses with moderate proximity to tax offices.

\newpage

\small\renewcommand{\baselinestretch}{1}\normalsize

\enlargethispage{\baselineskip}

\section*{Tables and Figures}

\singlespacing
\scriptsize
\begin{center}
\small{\textsc{Table 1: Pre-Treatment Characteristics}}\\
\scriptsize \centerline{\begin{tabular}{lccccccccccccccc}\\
\midrule
& \multicolumn{2}{c}{Control} &&\multicolumn{2}{c}{Treatment 1} && \multicolumn{2}{c}{Treatment 2} && \multicolumn{2}{c}{Treatment 3}  \\
\cmidrule(r){2-3} \cmidrule(r){5-6} \cmidrule(l){8-9} \cmidrule(l){11-12} 
&Mean &SD && Mean &SD & $p$-value & Mean &SD& $p$-value & Mean &SD& $p$-value \\
\midrule
~\\[-10pt]
\textbf{Variables used for stratification} &&&&&&&&&&&& \\
\ \ Annual turnover (million IDR) &1,120&1,672&&1,145&1,828&0.588&1,146&1,973&0.583&1,146&2,433&0.630\\
\ \ Age of business (years) &12.0&9.0&&11.9&9.3&0.693&11.9&8.8&0.489&11.9&8.8&0.487\\
\ \ Distance to tax office (kms) &6.2&8.8&&6.2&8.8&0.882&6.3&8.8&0.737&6.3&8.9&0.532\\
\ \ BTS density &1,746&1,599&&1,730&1,606&0.690&1,772&1,640&0.540&1,760&1,638&0.741\\
\ \ Jakarta &0.566&0.496&&0.563&0.496&0.804&0.555&0.497&0.420&0.559&0.497&0.604\\
\ \ South Sumatera &0.306&0.461&&0.309&0.462&0.835&0.316&0.465&0.422&0.308&0.462&0.907\\
\ \ West Nusa Tenggara &0.128&0.334&&0.128&0.335&0.936&0.128&0.335&0.932&0.133&0.340&0.546\\
\ \ Sector (1 Service; 0 Trade and industry) &0.506&0.500&&0.512&0.500&0.652&0.511&0.500&0.699&0.518&0.500&0.357\\
\ \ Type (1 Business only; 0 Diverse income) &0.382&0.486&&0.387&0.487&0.665&0.375&0.484&0.587&0.392&0.488&0.418\\
\ \ Tax return group (1 Individual; 0 Firm) &0.790&0.408&&0.784&0.412&0.595&0.797&0.403&0.503&0.782&0.413&0.492\\
\textbf{Additional variables} &&&&&&&&&&&& \\
\ \ Received penalty before &0.271&0.444&&0.281&0.449&0.386&0.274&0.446&0.766&0.297&0.457&0.024\\
\ \ Did not file more than 6 months &0.543&0.498&&0.547&0.498&0.738&0.543&0.498&0.992&0.565&0.496&0.088\\
\emph{N} &2,960&&&2,975&&&2,974&&&3,031&&\\

~\\[-10pt]
\midrule
\end{tabular}}
\scriptsize
\centerline{\parbox{19cm}{\emph{Note:} Treatment 1: Enforcement; Treatment~2: Information; Treatment 3: Use of tax revenue. $p$-values (based on robust standard errors) refer to the comparison of means between treatment and control groups.}}
\end{center}
\normalsize
\doublespacing

\newpage

\singlespacing
\scriptsize
\begin{center}
\small{\textsc{Table 2: Treatment Effects}}\\
\scriptsize \centerline{\begin{tabular}{lcccccccc}\\
\midrule
 & & T1 & T2 & T3 &  &  &  \\[.5mm]
&Control & vs.  & vs. & vs. &  \multicolumn{3}{c}{$p$-values}  \\
 \cline{6-8} \\[-2.5mm]
& Mean & Control  & Control  & Control  & T1$=$T2 & T1$=$T3 & T2$=$T3  \\
\midrule
~\\[-10pt]
\textbf{Filing} &&&\\[5pt]
Filed tax return &0.5706& 0.2396***&0.2325***&0.0869***&0.482&0.000&0.000\\
&  &(0.0099)&(0.0099)&(0.0102)&&&\\[5pt]
Filed non-zero return &0.1564& 0.3979***&0.0800***&0.1214***&0.000&0.000&0.000\\
&  &(0.0109)&(0.0100)&(0.0101)&&&\\[5pt]
\hline \\[-2mm]
\textbf{Payment} &&&\\[5pt]
Increased payment &0.1599& 0.4258***&0.0917***&0.1262***&0.000&0.000&0.001\\
&  &(0.0105)&(0.0099)&(0.0100)&&&\\[5pt]
Additional amount paid (IDR) &   10,342&    39,379***&    5,565***&    3,812***&0.000&0.000&0.008\\
&  &(931.26)&(670.18)&(591.30)&&&\\[5pt]

~\\[-15pt]
\midrule
\vspace{-0.3cm}
\end{tabular}}
\scriptsize
\centerline{\parbox{14.5cm}{\emph{Note:} T1: Enforcement; T2: Information; T3: Use of tax revenue. The pre-treatment characteristics presented in Table~1 are used as covariates. Standard errors (presented in parentheses) were clustered to account for repeated observations over time. \\[2mm]
** $p<0.05$, *** $p<0.01.$}}
\end{center}
\normalsize
\doublespacing

\newpage

\singlespacing
\scriptsize
\begin{center}
\small{\textsc{Table 3: Treatment Effects by Region}}\\
\scriptsize \centerline{\begin{tabular}{lccccccccccccc}\\
\midrule
&\multicolumn{3}{c}{\textbf{Jakarta}} && \multicolumn{3}{c}{\textbf{South Sumatra}} && \multicolumn{3}{c}{\textbf{West Nusa Tenggara}} \\
\cline{2-4} \cline{6-8} \cline{10-12} \\[-2mm]
 & T1 & T2 & T3 && T1 & T2 & T3  && T1 & T2 & T3&\\
 &vs. &vs. &vs. &&vs. &vs. &vs. &&vs. &vs. &vs. \\
 &Control &Control &Control &&Control &Control &Control &&Control &Control &Control \\ 
\midrule
~\\[-10pt]
Filed tax return &-0.1005***&-0.0585***&-0.0695***&&0.7663***&0.6721***&0.3148***&&0.4760***&0.4475***&0.2338***\\
 &(0.0111)&(0.0107)&(0.0108)&&(0.0147)&(0.0169)&(0.0197)&&(0.0263)&(0.0266)&(0.0254)\\[5pt]
Filed non-zero return &0.3080***&0.0055&0.1162***&&0.6174***&0.2158***&0.1648***&&0.2661***&0.0779***&0.0433***\\
 &(0.0161)&(0.0153)&(0.0159)&&(0.0164)&(0.0143)&(0.0132)&&(0.0227)&(0.0151)&(0.0121)\\[5pt]
Increased payment &0.3347***&0.0172&0.1196***&&0.6383***&0.2266***&0.1722***&&0.3170***&0.0925***&0.0480***\\
 &(0.0155)&(0.0152)&(0.0158)&&(0.0155)&(0.0142)&(0.0131)&&(0.0217)&(0.0151)&(0.0123)\\[5pt]
Additional amount  &   33,874***&      783&    1,860&&   55,421***&   14,547***&    8,179***&&   25,090***&    4,882***&    1,935***\\
\ \ paid (IDR) &(1324.00)&(1027.05)&(947.46)&&(1518.12)&(973.77)&(719.50)&&(2090.41)&(981.20)&(683.36)\\[5pt]

~\\[-15pt]
\midrule
\vspace{-0.3cm}
\end{tabular}}
\scriptsize
\centerline{\parbox{19.5cm}{\emph{Note:} T1: Enforcement; T2: Information; T3: Use of tax revenue. The pre-treatment characteristics presented in Table~1 are used as covariates. Standard errors (presented in parentheses) were clustered to account for repeated observations over time. \\[2mm]
** $p<0.05$, *** $p<0.01.$}}
\end{center}
\normalsize
\doublespacing

\newpage

\singlespacing
\scriptsize
\begin{center}
\small{\textsc{Table 4: Control Group Means by Region and Distance to Local Tax Office}} \\
\scriptsize \centerline{\begin{tabular}{lcccccccccccc}\\
\midrule
& && \multicolumn{10}{c}{\textsc{Distance to Local Tax Office (Deciles)}}  \\[1mm]
\cline{4-13}\\[-2mm]
 & All && 1 &2  & 3 &4 & 5 &6  &7 &8 & 9 &10  \\ 
\midrule
~\\[-10pt]
\textbf{Jakarta} &&\\[3pt]
 \ \ Median distance (kms) &&&0.5&0.6&0.7&0.8&0.9&1.0&1.1&1.2&1.3&1.4\\[3pt]
\ \ Filed tax return &0.923&&1.000&1.000&1.000&1.000&1.000&1.000&1.000&1.000&0.856&0.403\\
\ \ Filed non-zero return &0.267&&0.988&0.561&0.240&0.202&0.151&0.137&0.086&0.107&0.138&0.097\\
\ \ Increased payment &0.271&&0.987&0.564&0.243&0.209&0.153&0.140&0.090&0.111&0.139&0.102\\
\ \ Additional amount paid (IDR) &   17,588&&   64,310&   36,491&   15,914&   13,451&   10,168&    8,850&    5,676&    7,409&    8,936&    6,759\\
\emph{N} & 20,100&&  1,980&  1,860&  2,196&  2,136&  1,824&  2,016&  1,944&  2,028&  2,004&  2,112\\
\midrule
\textbf{South Sumatra} &&\\[3pt]
 \ \ Median distance (kms) &&&6.0&6.3&6.4&6.5&6.7&6.8&6.9&7.0&7.1&7.5\\[3pt]
\ \ Filed tax return &0.137&&0.392&0.232&0.277&0.128&0.040&0.055&0.026&0.034&0.093&0.045\\
\ \ Filed non-zero return &0.014&&0.010&0.012&0.010&0.021&0.027&0.018&0.013&0.011&0.010&0.011\\
\ \ Increased payment &0.018&&0.014&0.015&0.017&0.021&0.029&0.021&0.019&0.014&0.015&0.017\\
\ \ Additional amount paid (IDR) &    1,065&&      832&      763&      803&    1,402&    1,823&    1,359&      981&      893&    1,024&      826\\
\emph{N} & 10,884&&  1,164&    984&  1,212&  1,128&    900&  1,308&    912&  1,056&  1,164&  1,056\\
\midrule
\textbf{West Nusa Tenggara} &&\\[3pt]
 \ \ Median distance (kms) &&&22.8&27.5&27.6&27.8&27.9&28.1&28.2&28.4&30.4&31.0\\[3pt]
\ \ Filed tax return &0.050&&0.098&0.059&0.025&0.032&0.026&0.050&0.049&0.083&0.073&0.000\\
\ \ Filed non-zero return &0.005&&0.024&0.000&0.000&0.000&0.000&0.000&0.000&0.028&0.000&0.000\\
\ \ Increased payment &0.010&&0.033&0.005&0.000&0.005&0.000&0.002&0.000&0.037&0.008&0.007\\
\ \ Additional amount paid (IDR) &      494&&    1,791&       25&        0&       24&        0&       15&        0&    2,424&      284&      340\\
\emph{N} &  4,536&&    492&    408&    480&    372&    456&    480&    492&    432&    492&    432\\

~\\[-10pt]
\midrule
\vspace{-0.3cm}
\end{tabular}}
\end{center}
\normalsize
\doublespacing

\newpage

\small\renewcommand{\baselinestretch}{1}\normalsize

\begin{center}
\small{\textsc{Figure 1: Treatment Effects on Filing Behavior by Distance to Local Tax Office (Deciles)}}\\[.5cm]
\textbf{Jakarta} \\
\begin{figure}[!h]
  \centering
  \begin{minipage}[b]{0.45\textwidth}
    \includegraphics[width=\textwidth]{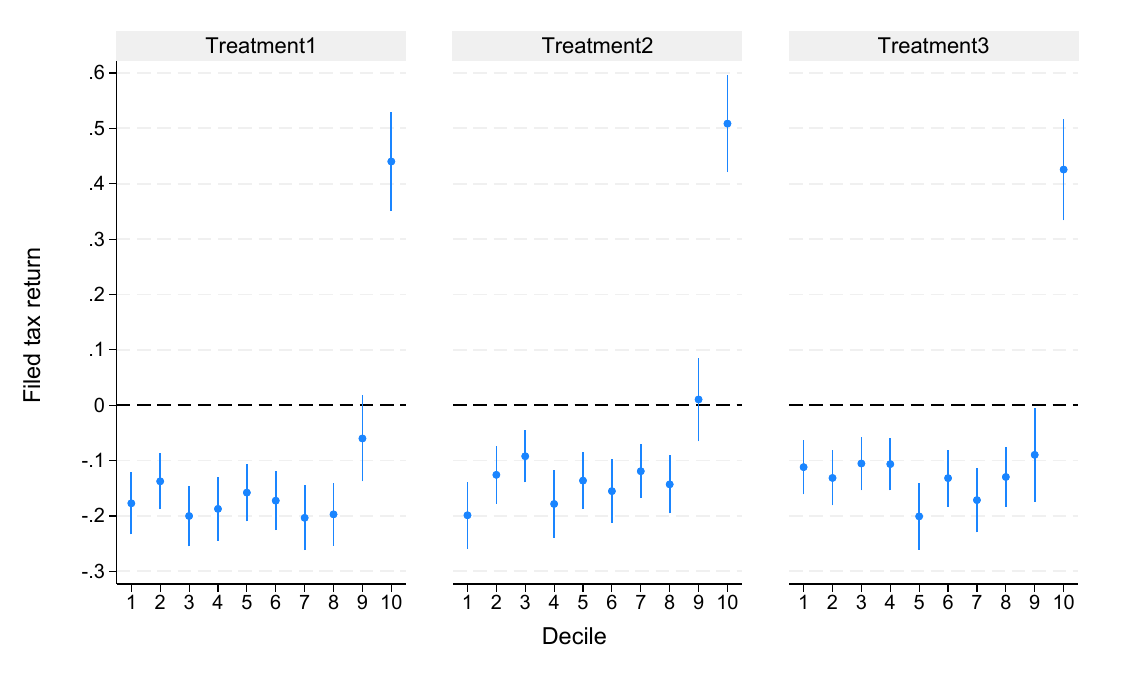} 
  \end{minipage}
  \begin{minipage}[b]{0.45\textwidth}
    \includegraphics[width=\textwidth]{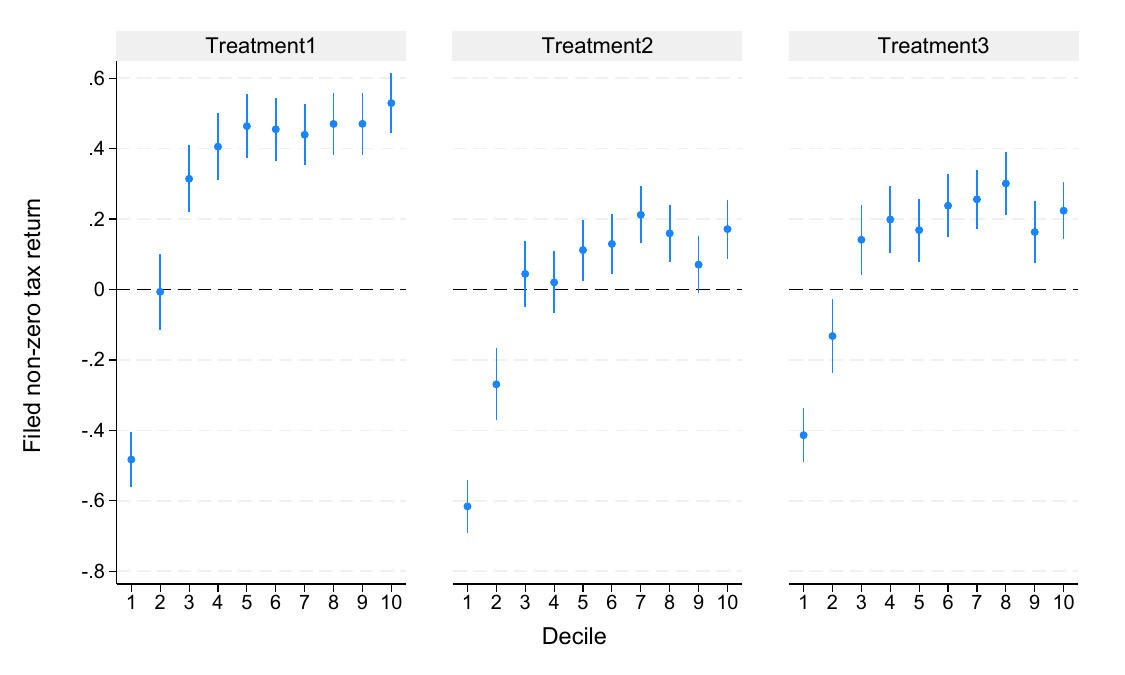}
  \end{minipage}
\end{figure}
\vspace{-5mm}
\textbf{South Sumatra} \\
\begin{figure}[!h]
  \centering
  \begin{minipage}[b]{0.45\textwidth}
    \includegraphics[width=\textwidth]{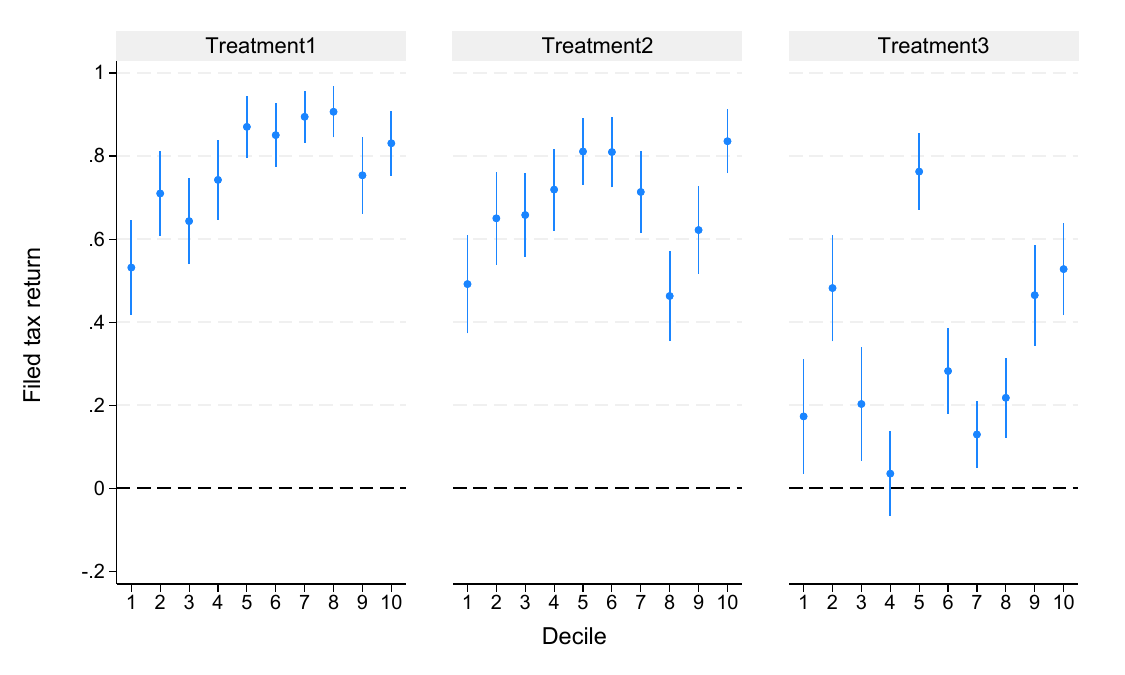} 
  \end{minipage}
  \begin{minipage}[b]{0.45\textwidth}
    \includegraphics[width=\textwidth]{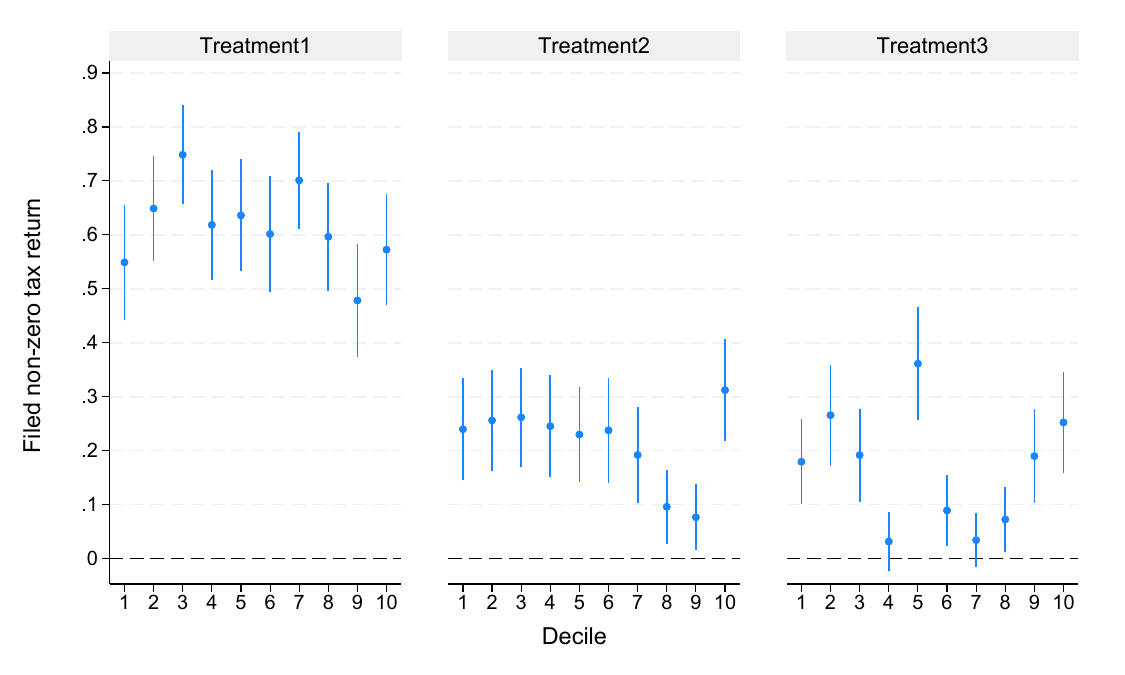}
  \end{minipage}
\end{figure}
\vspace{-5mm}
\textbf{West Nusa Tenggara} \\
\begin{figure}[!h]
  \centering
  \begin{minipage}[b]{0.45\textwidth}
    \includegraphics[width=\textwidth]{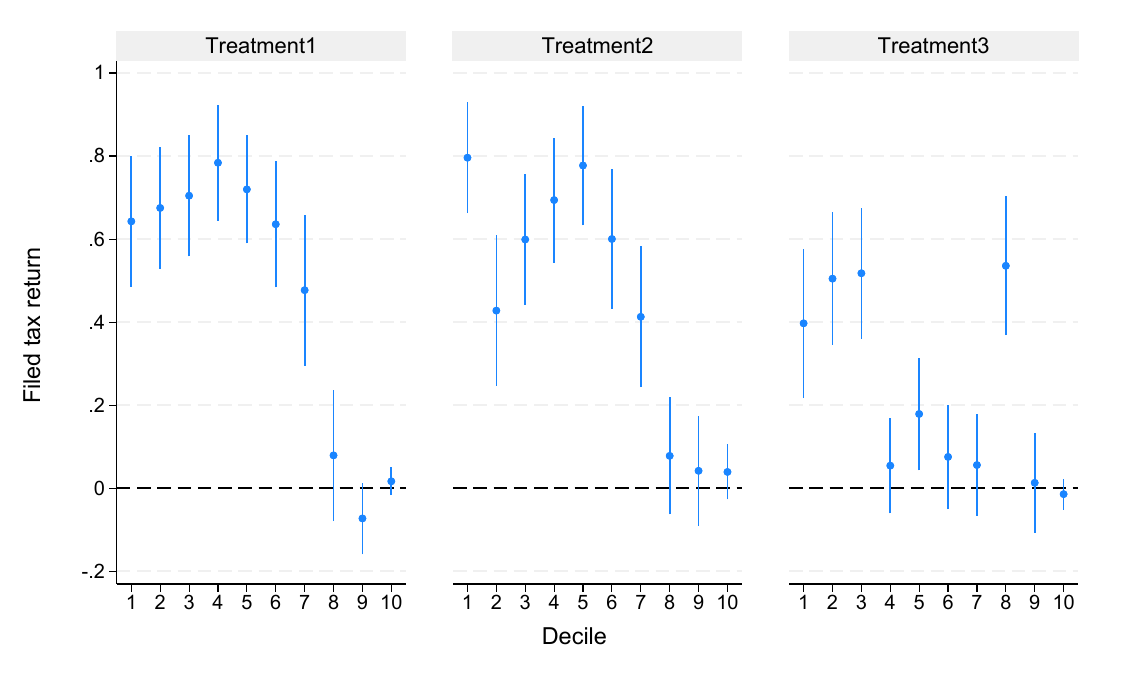} 
  \end{minipage}
  \begin{minipage}[b]{0.45\textwidth}
    \includegraphics[width=\textwidth]{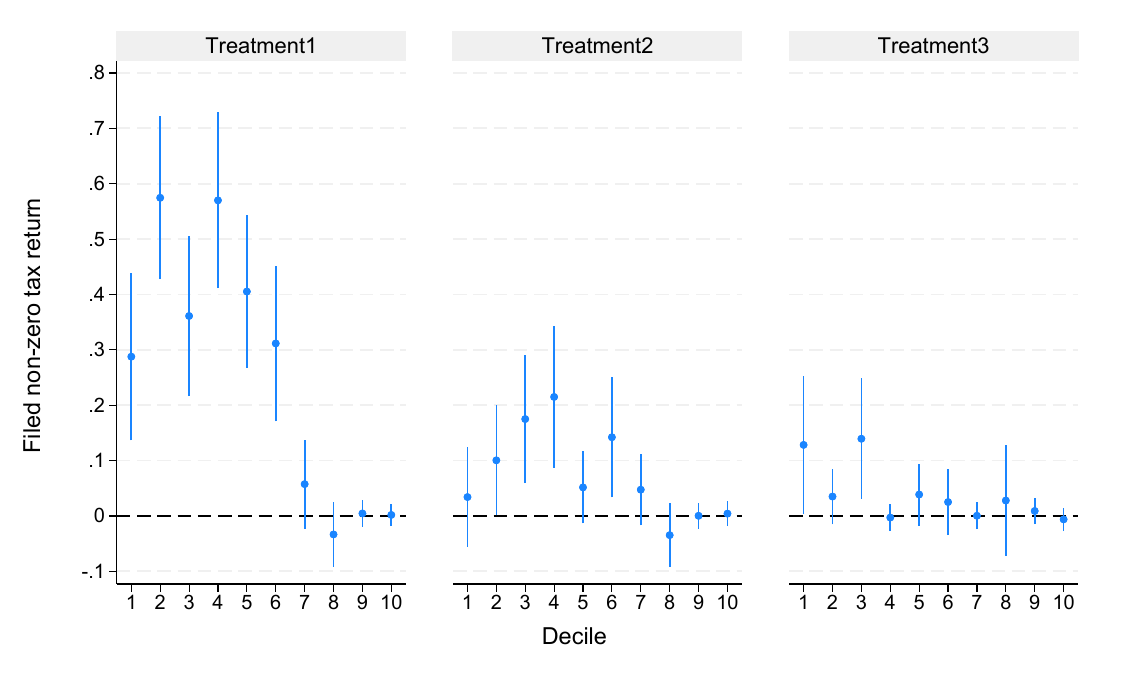}
  \end{minipage}
\end{figure}
\scriptsize
\centerline{\parbox{13cm}{\emph{Note:} The pre-treatment characteristics presented in Table~1 are used as covariates. Standard errors were clustered to account for repeated observations over time.}}
\end{center}

\newpage

\small\renewcommand{\baselinestretch}{1}\normalsize

\begin{center}
\small{\textsc{Figure 2: Treatment Effects on Payment Behavior by Distance to Local Tax Office (Deciles)}}\\[.5cm]
\textbf{Jakarta} \\
\begin{figure}[!h]
  \centering
  \begin{minipage}[b]{0.45\textwidth}
    \includegraphics[width=\textwidth]{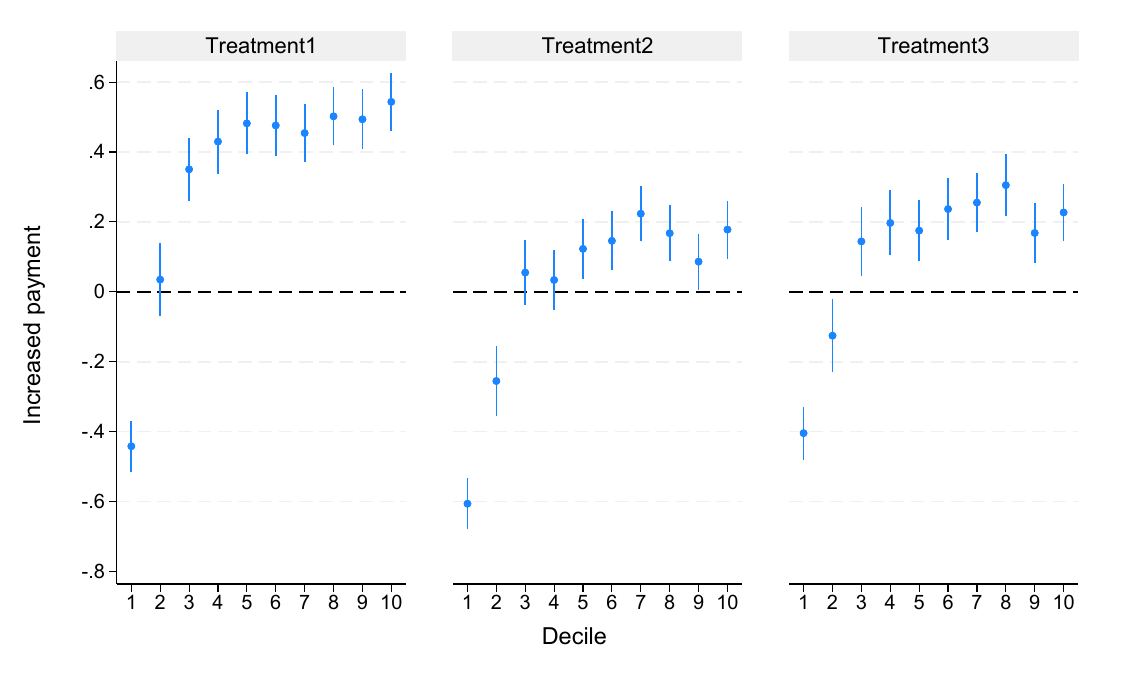} 
  \end{minipage}
  \begin{minipage}[b]{0.45\textwidth}
    \includegraphics[width=\textwidth]{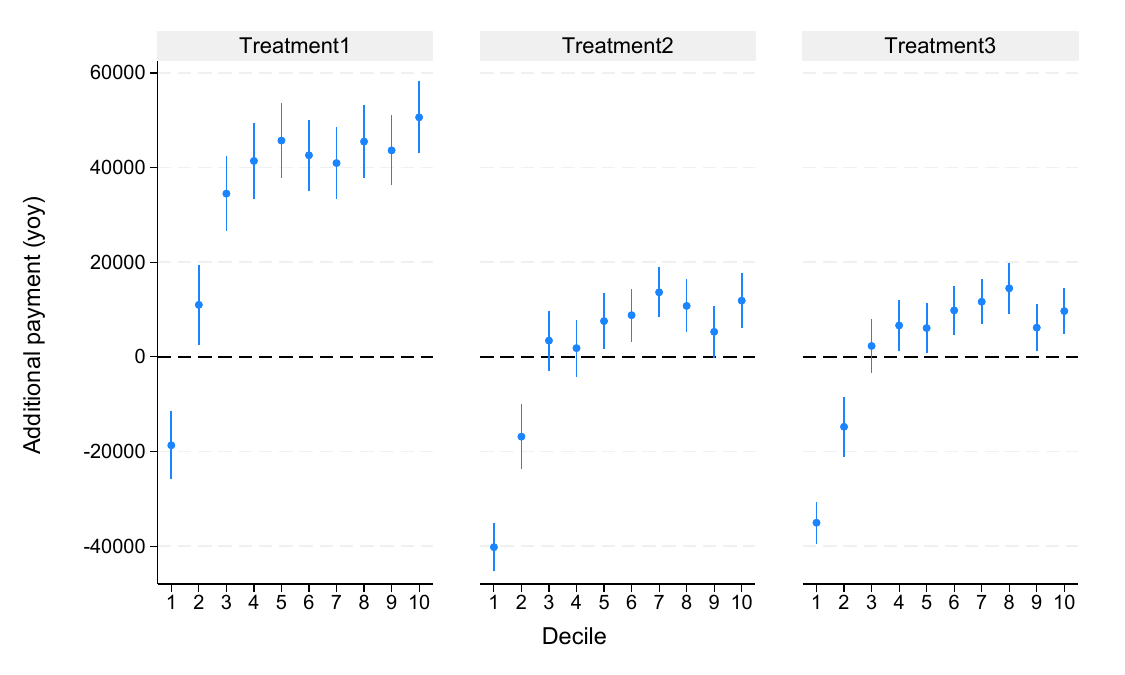}
  \end{minipage}
\end{figure}
\vspace{-5mm}
\textbf{South Sumatra} \\
\begin{figure}[!h]
  \centering
  \begin{minipage}[b]{0.45\textwidth}
    \includegraphics[width=\textwidth]{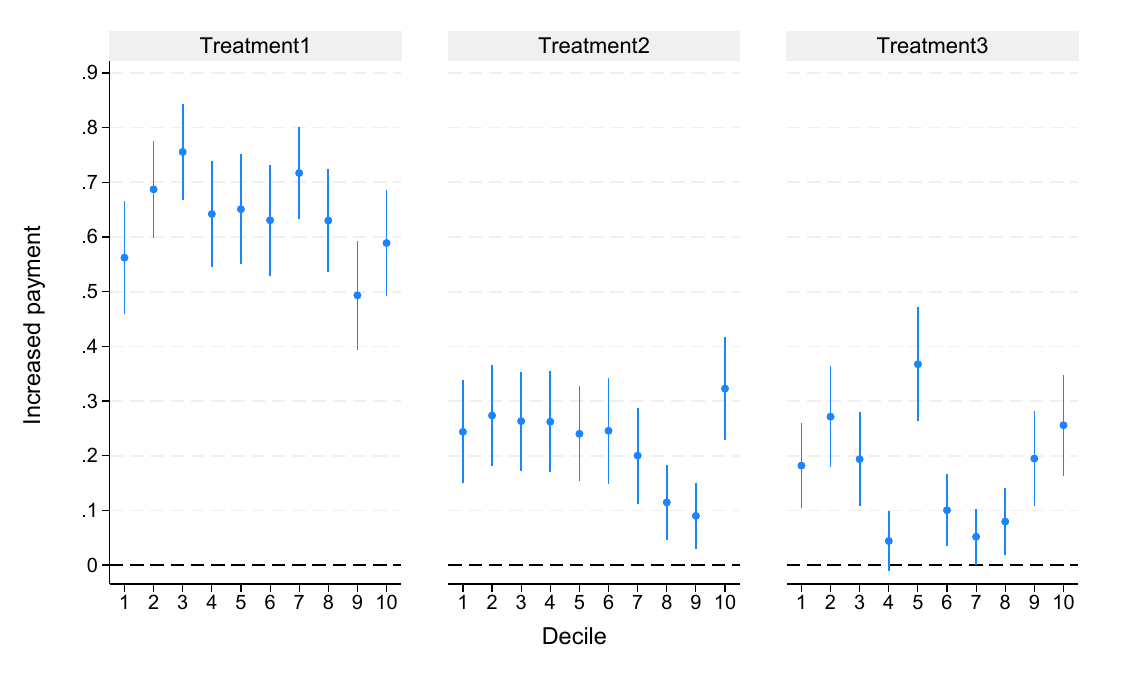} 
  \end{minipage}
  \begin{minipage}[b]{0.45\textwidth}
    \includegraphics[width=\textwidth]{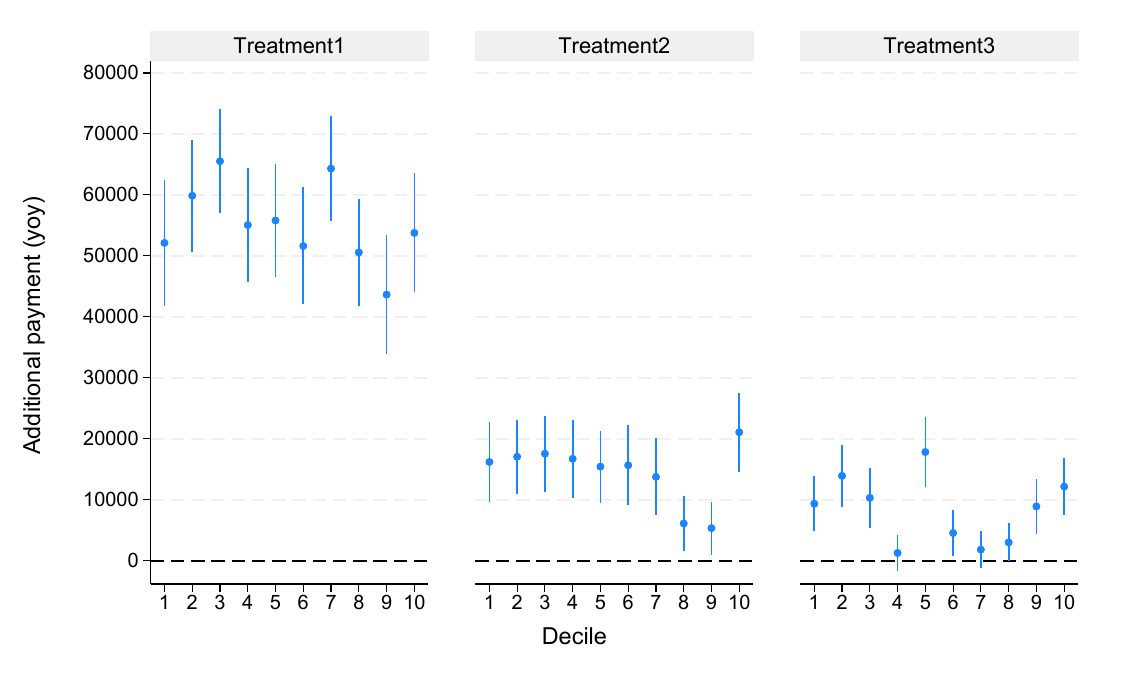}
  \end{minipage}
\end{figure}
\vspace{-5mm}
\textbf{West Nusa Tenggara} \\
\begin{figure}[!h]
  \centering
  \begin{minipage}[b]{0.45\textwidth}
    \includegraphics[width=\textwidth]{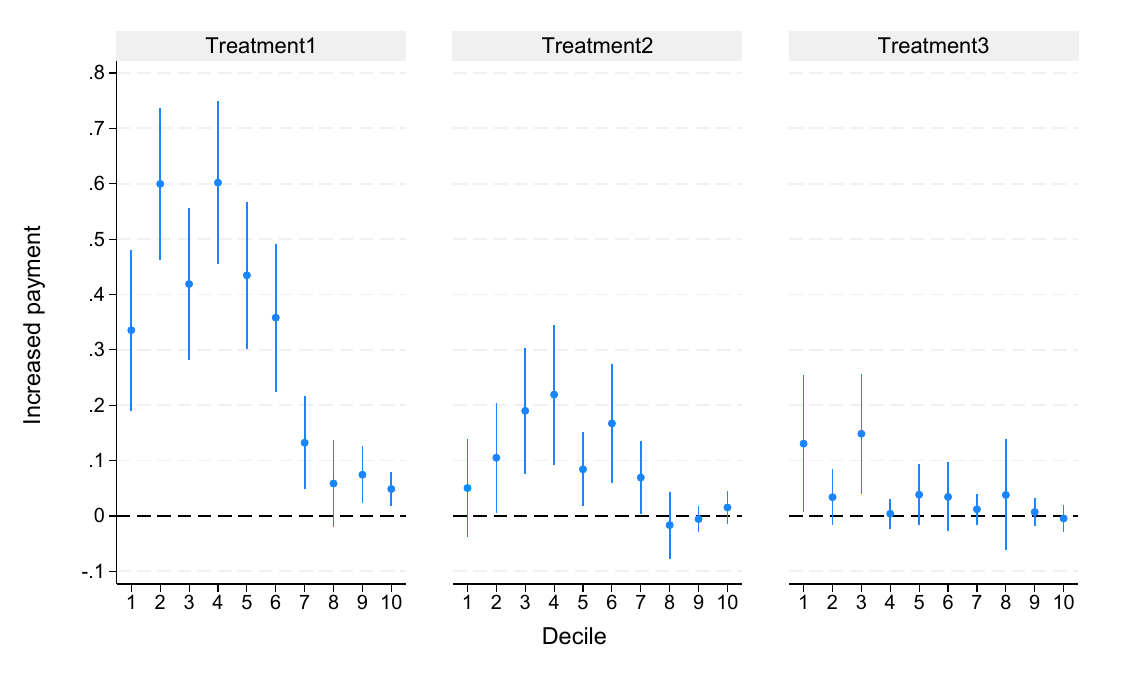} 
  \end{minipage}
  \begin{minipage}[b]{0.45\textwidth}
    \includegraphics[width=\textwidth]{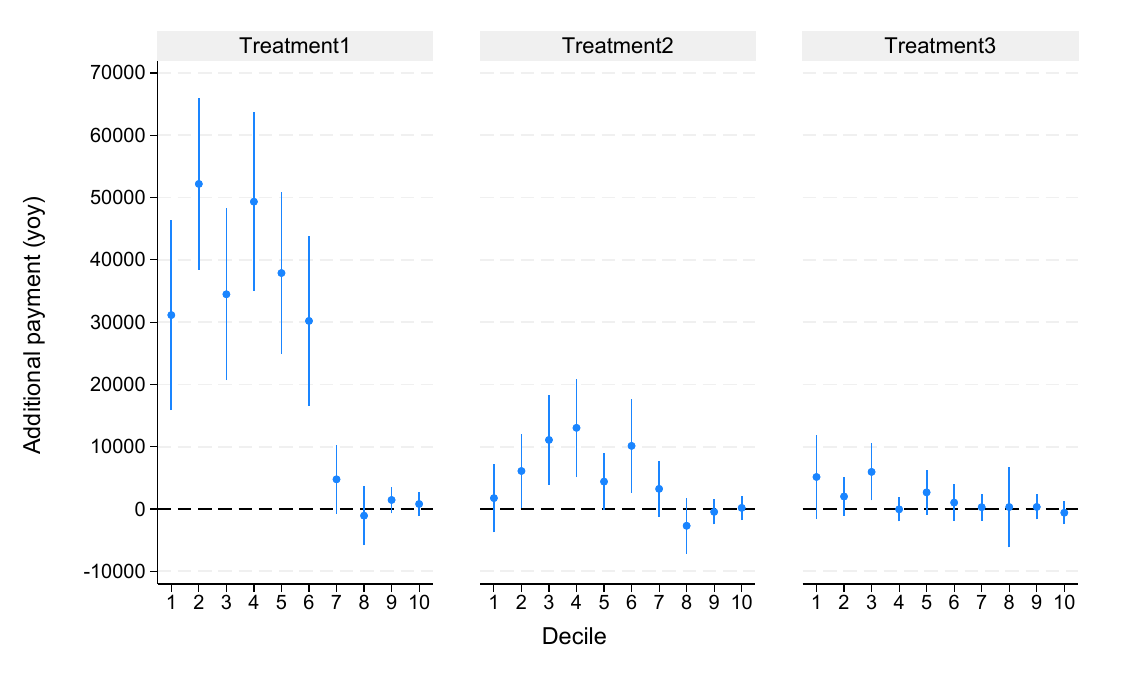}
  \end{minipage}
\end{figure}
\scriptsize
\centerline{\parbox{13cm}{\emph{Note:} The pre-treatment characteristics presented in Table~1 are used as covariates. Standard errors were clustered to account for repeated observations over time.}}
\end{center}

\newpage

\begin{center}
	\small{\textsc{Figure 3: Treatment effects on Payment Behavior by Sub-District in North Jakarta}}\\[0cm]
	\includegraphics[width=16cm]{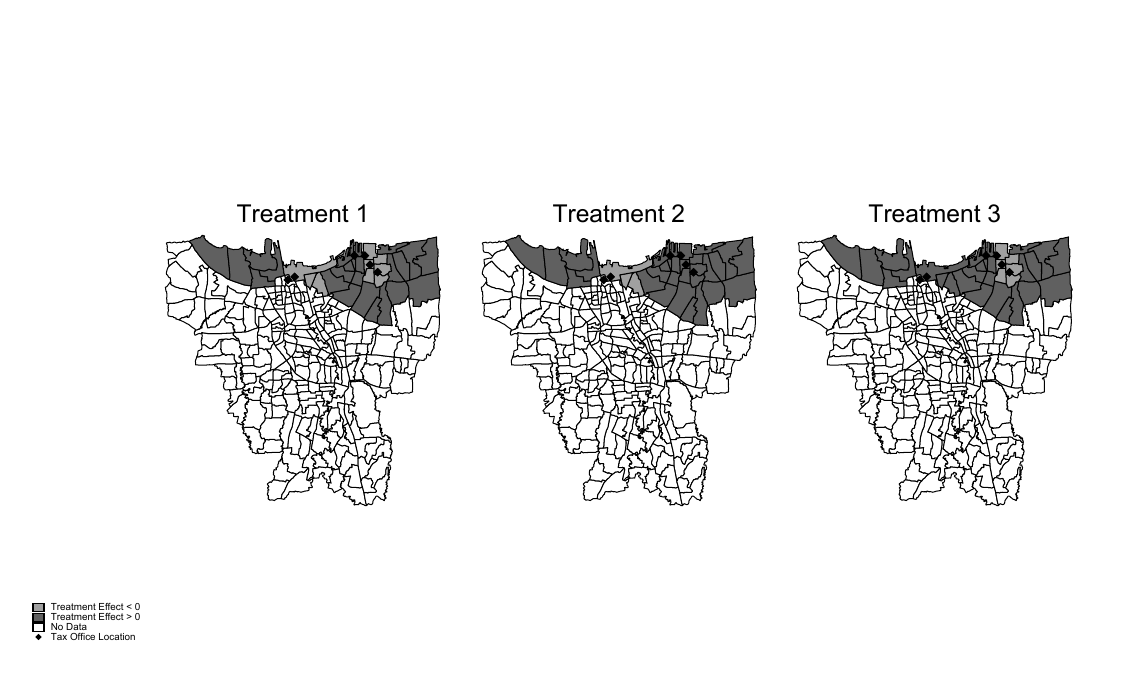}
\scriptsize
\centerline{\parbox{14.5cm}{\emph{Note:} The maps are constructed using geospatial data provided by Indonesia's Statistics Bureau, merged with subdistrict-level data from the experimental design. The outcome variable represents the additional amount of tax paid. Tax office locations are accurately mapped using precise longitude and latitude coordinates.}}
\end{center}

\newpage

\singlespacing
\scriptsize
\begin{center}
\textsc{\small{Table 5: Cost-Benefit Analysis}}\\[.5cm]
\centering
\begin{tabular}{lccc}
\toprule
& T1 (Enforcement) & T2 (Information) & T3 (Use of tax revenue) \\
\midrule
\textbf{Amount of tax paid (IDR)}  \\
\ \ Monthly tax paid & 39,379 & 5,565 & 3,812 \\
\ \ Annual tax paid (present value)  & 455,933 & 64,432 & 44,136 \\\\
\textbf{Costs (IDR)} \\
\ \ Printing: cost per letter & 800 & 1,000 & 800 \\
\ \ Printing: cost per flyer & 0 & 3,000 & 3,000 \\
\ \ Postage: cost per letter & 14,402 & 14,402 & 14,402 \\
\ \ Total costs & 15,202 & 18,402 & 18,202 \\ \\
\textbf{Benefit-cost ratio} & 29.99 & 3.50 & 2.42 \\[-3mm]

 \\
\bottomrule
\vspace{-2mm}
\end{tabular}
\centerline{\parbox{13cm}{\emph{Note:} The present value calculations are based on a 6.66 percent real rate of interest, in line with \citet{worldbank2020_rct}. Postage costs per letter are based on the price of sending a standard letter using Indonesia Post. For more information, visit the official Indonesia Post website. }}
\end{center}

\normalsize
\doublespacing

\newpage

\renewcommand{\baselinestretch}{1} \normalsize

\bibliographystyle{aer}
\bibliography{DGT}
\addcontentsline{toc}{section}{References} 

\newpage

\section*{Appendix A: Location of trial sample}

\begin{center}
	\includegraphics[width=16cm]{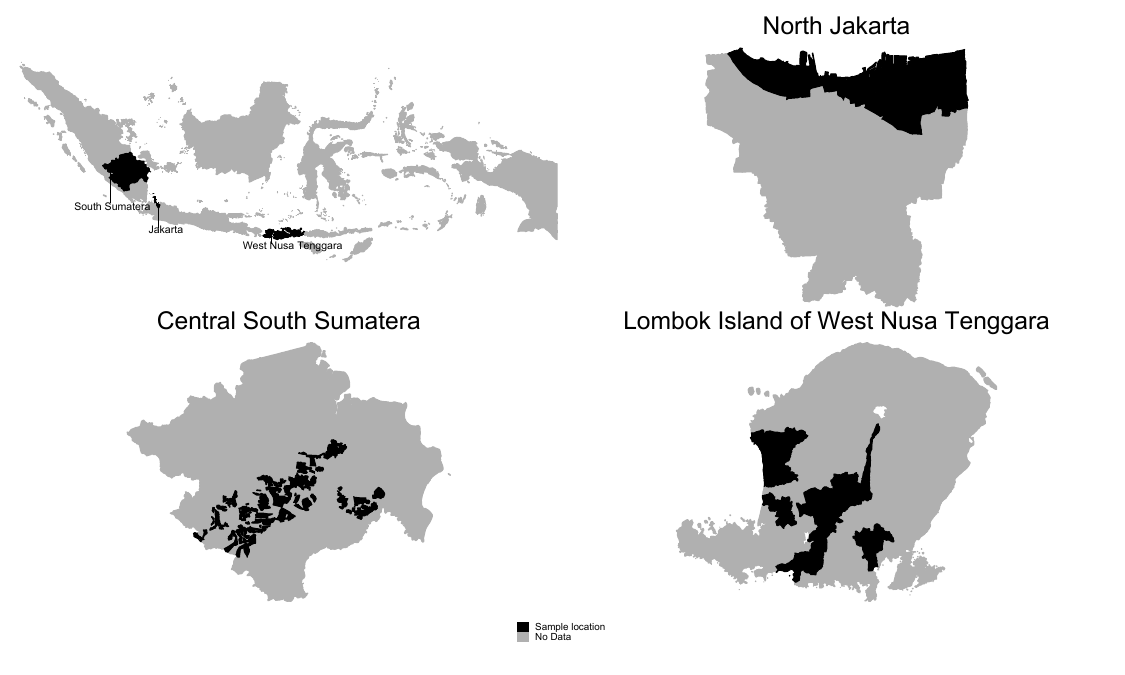}
\scriptsize
\centerline{\parbox{15cm}{\emph{Note:} The maps are based on geospatial data from Indonesia's Statistics Bureau that have been merged with subdistrict-level data from the experimental design.}}
\end{center}

\newpage

\section*{Appendix B: Treatment letters}

\vspace{5mm}

\begin{center}
\small{\textsc{Treatment 1: Enforcement}}\\[.1cm]
\includegraphics[width=16cm]{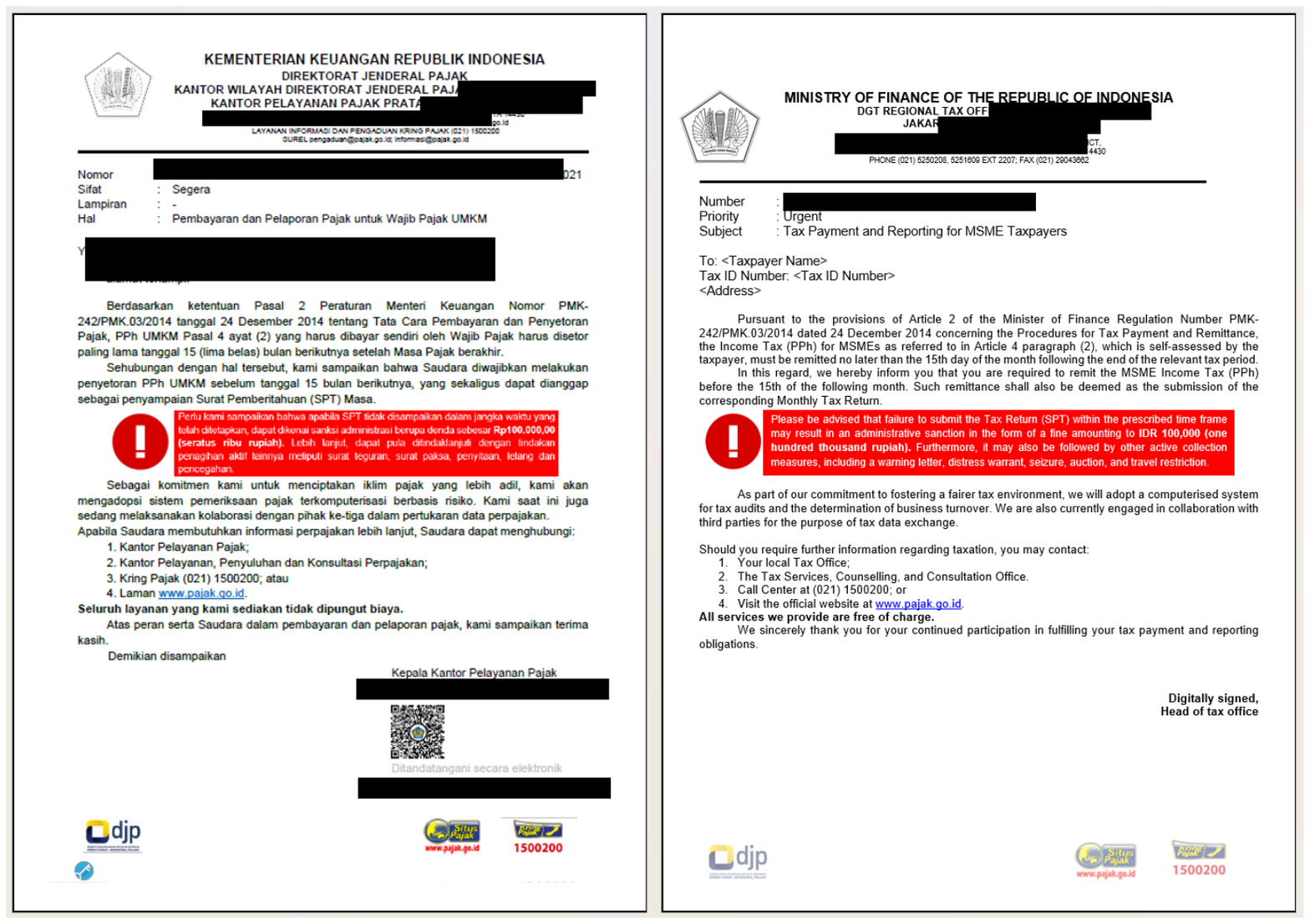}
\scriptsize
\centerline{\parbox{14cm}{\emph{Note:} This treatment letter highlights the consequences of failing to submit the tax return (SPT) on time. It outlines potential penalties, including monetary fines and enforcement actions such as warning letters, forced collection orders, prevention measures, and auctions. The letter follows official correspondence procedures of the DGT and is delivered to the taxpayer's registered address as listed in the DGT database. To verify the letter's authenticity, recipients can scan the QR code linked to the electronic signature. This redirects to a digital copy of the letter stored in the Ministry of Finance's official correspondence system.}}
\end{center}

\newpage

\begin{center}
\small{\textsc{Treatment 2: Information}}\\[.1cm]
\includegraphics[width=16cm]{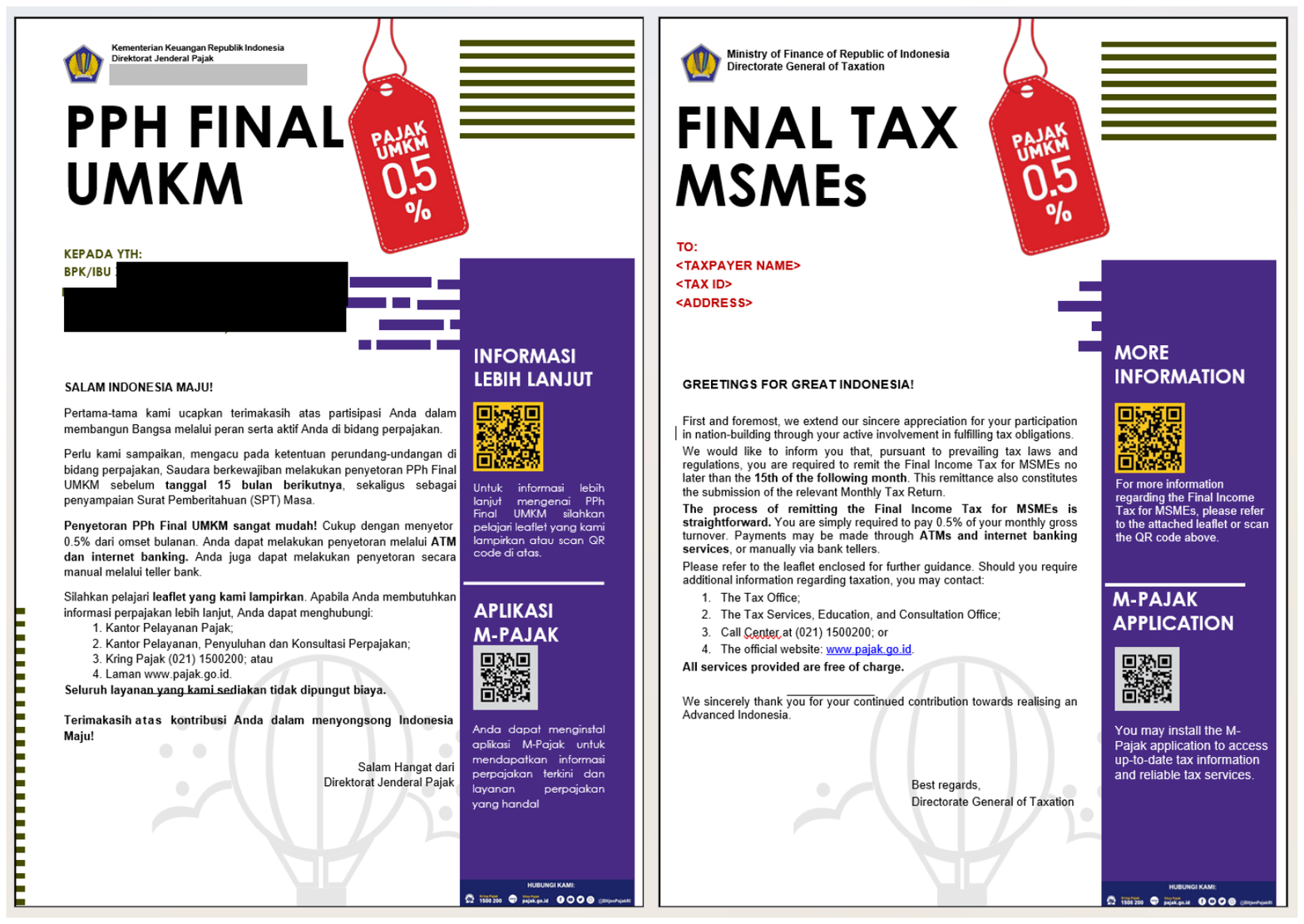}
\scriptsize
\centerline{\parbox{14cm}{\emph{Note:} This treatment letter serves as a cover letter accompanying instructions on how to calculate, pay, and file the tax return (SPT) for MSMEs. It emphasizes the submission deadline and underscores that fulfilling MSME tax obligations is a simple and straightforward process. The letter follows official correspondence procedures of the DGT and is delivered to the taxpayer's registered address as listed in the DGT database. To verify the letter's authenticity, recipients can scan the QR code linked to the electronic signature. This redirects to a digital copy of the letter stored in the Ministry of Finance's official correspondence system.}}
\end{center}

\newpage

\begin{center}
\small{\textsc{Treatment 2: Flyer (Page 1)}}\\[.1cm]
\begin{textblock*}{2cm}(4.2cm, 9cm) 
	\rotatebox{90}{\footnotesize Original}
\end{textblock*}

\begin{textblock*}{2cm}(4.2cm, 17cm) 
	\rotatebox{90}{\footnotesize Translation}
\end{textblock*}
\includegraphics[width=12cm]{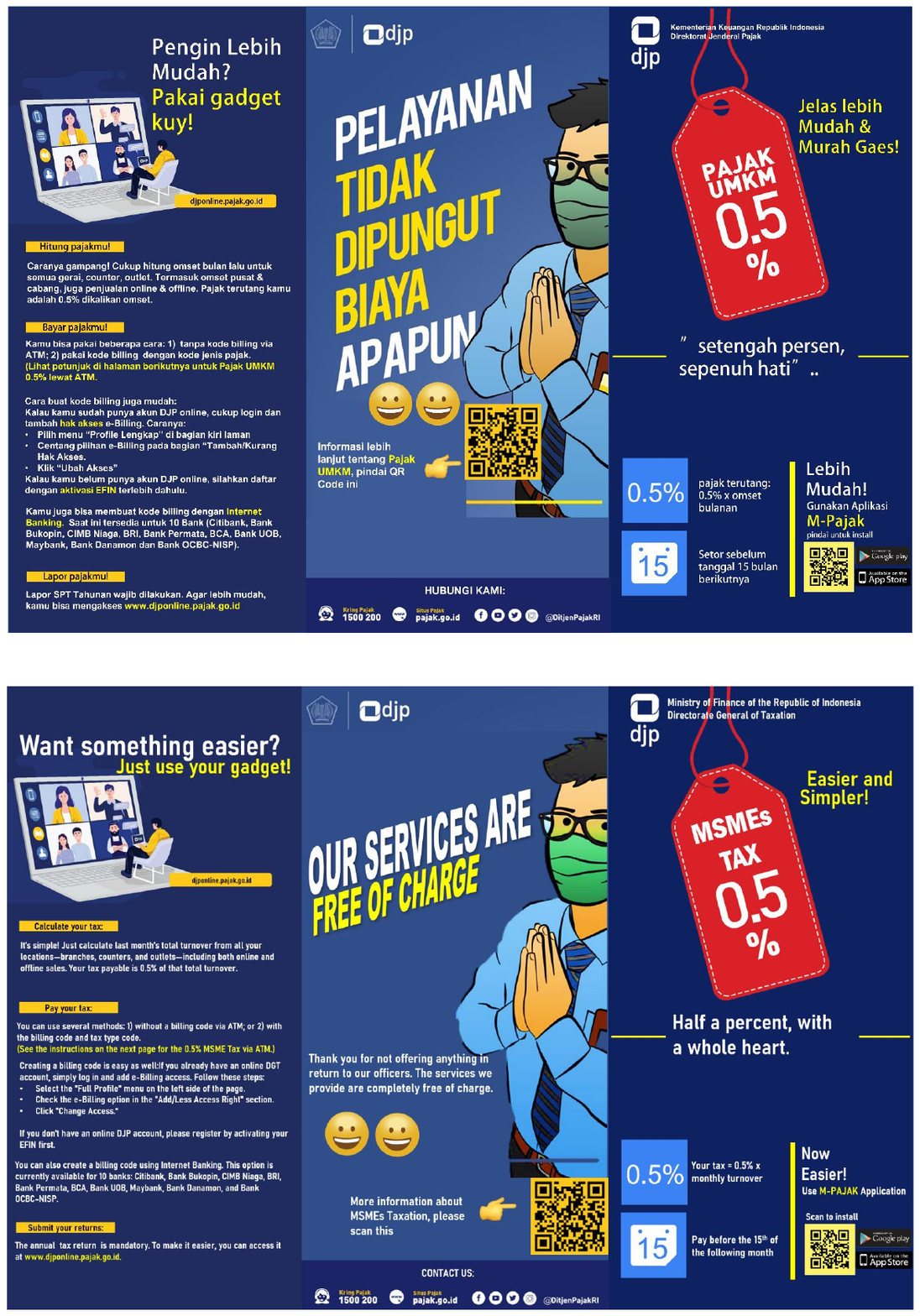}
\scriptsize
\centerline{\parbox{16cm}{\emph{Note:} This guide provides a clear overview of MSME tax compliance, organized into three main sections: \textbf{Left Panel:} Tax compliance for MSMEs is made easy through the use of digital tools. To calculate your monthly tax obligation, add up revenue from all sources~-- including stores, counters, outlets, headquarters, branches, and online or offline sales~-- from the previous month. The applicable tax rate is~0.5\% of this total. Payments can be made via ATM without a billing code or through DJP Online with a billing code. To generate a billing code, log in to DJP Online, activate e-Billing through the ``Full Profile'' menu, and update your access rights. If you don’t yet have an account, registration requires activation of your EFIN. Billing codes can also be created through internet banking, supported by~10 banks including Citibank, BRI, and BCA. Filing your annual tax return is mandatory and can be completed online at: \url{www.djponline.pajak.go.id}. \textbf{Center Panel:} All tax-related services are completely free of charge. For more information, scan the QR code on the guide. \textbf{Right Panel:} MSME tax is straightforward and affordable~-- just 0.5\% of your total monthly revenue. Payments are due by the 15th of the following month. To streamline the process, download and use the M-Pajak mobile app.}}
\end{center}

\newpage

\begin{center}
\small{\textsc{Treatment 2: Flyer (Page 2)}}\\[.1cm]
\begin{textblock*}{2cm}(4.2cm, 9cm) 
	\rotatebox{90}{\footnotesize Original}
\end{textblock*}

\begin{textblock*}{2cm}(4.2cm, 17cm) 
	\rotatebox{90}{\footnotesize Translation}
\end{textblock*}
\includegraphics[width=12cm]{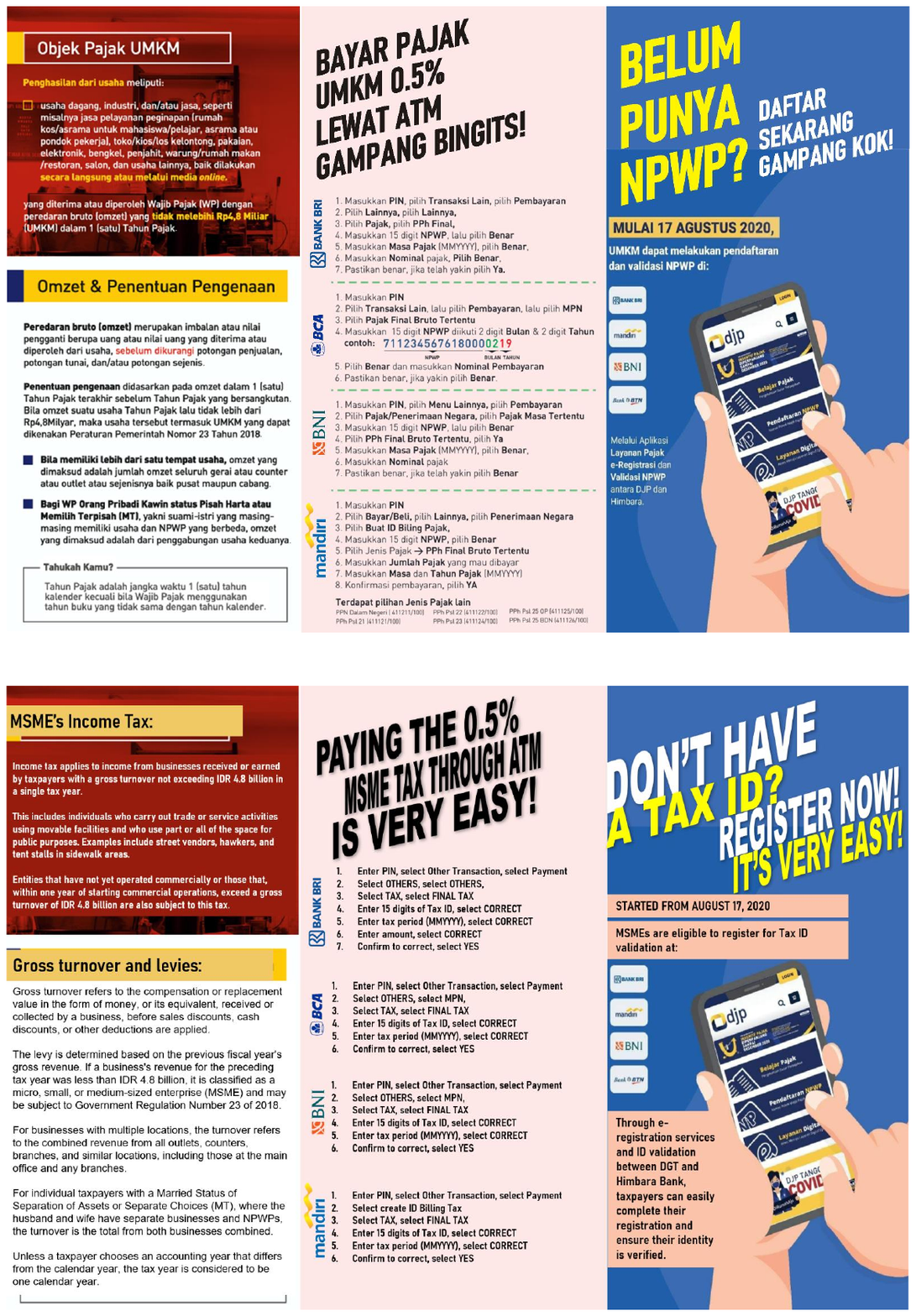}
\end{center}
\scriptsize
\centerline{\parbox{16cm}{\emph{Note:} This guide outlines the essentials of MSME tax compliance, presented in three clear sections: \textbf{Left Panel:} MSME tax applies to income earned from a wide range of business activities, including trade, manufacturing, and services. This covers operations such as accommodation services (e.g., student housing, dormitories, and worker lodgings), retail shops and kiosks, clothing and electronics stores, repair workshops, tailoring services, food stalls and restaurants, salons, and both online and offline businesses. The tax applies to businesses with gross annual revenue not exceeding IDR 4.8 billion. \textbf{Center Panel:} Paying MSME tax~-- just 0.5\% of gross monthly revenue~-- is simple, especially via ATM. Step-by-step payment instructions are available for each participating bank. These include how to navigate the ATM menu, select the tax type, enter your Taxpayer Identification Number (NPWP), and input the payment amount. A helpful list of tax type codes and deposit codes is also provided to make the process even smoother. \textbf{Right Panel:} Don’t Have an NPWP yet? Register now~-- it’s easy!
Getting your NPWP is quick and easy. As of 17 August 2020, MSMEs can register for and validate their NPWP directly through partner bank applications or by using the M-Pajak app.}}

\newpage

\begin{center}
\small{\textsc{Treatment 3: Use of taxpayer money}}\\[.1cm]
\includegraphics[width=16cm]{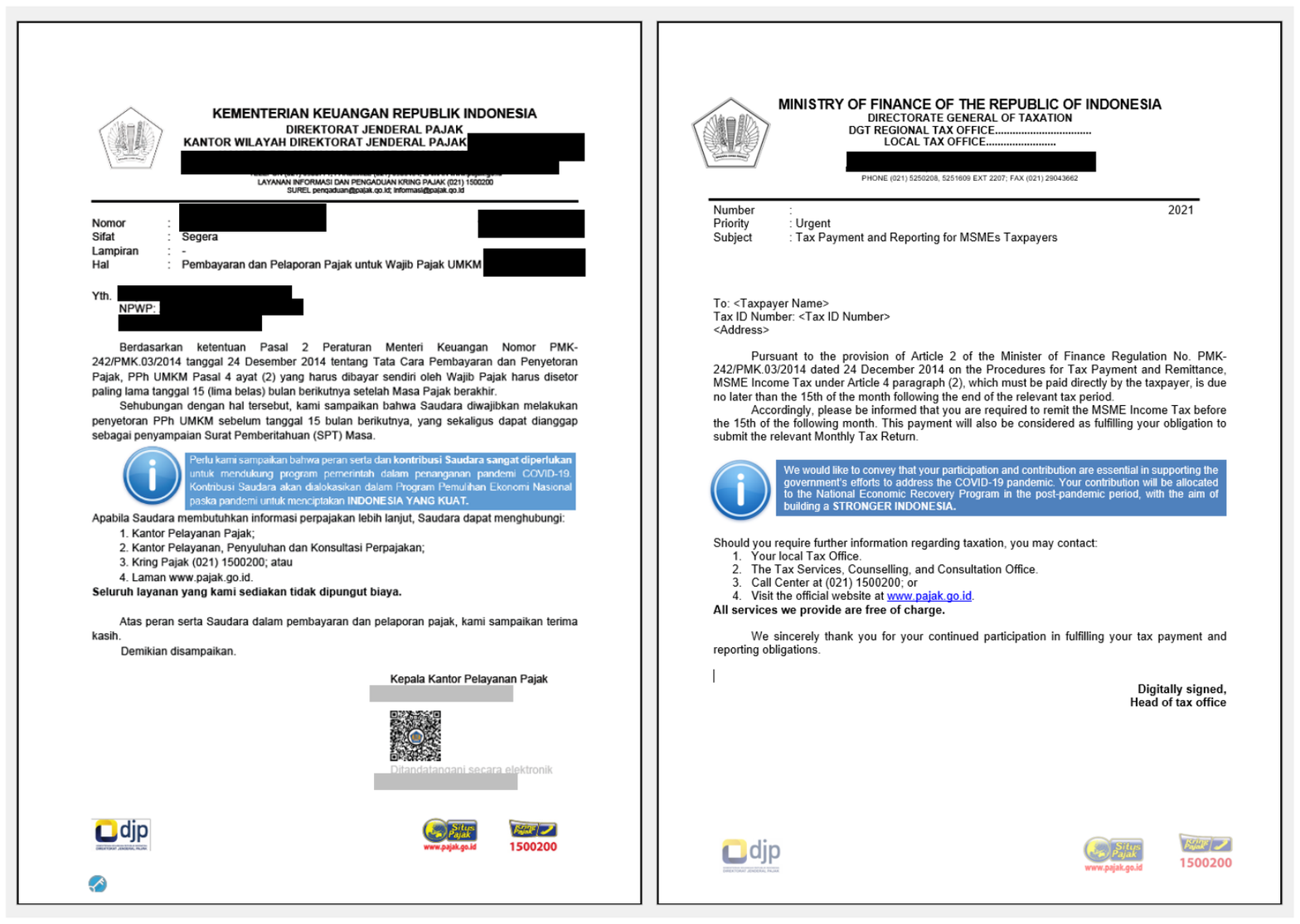}
\scriptsize
\centerline{\parbox{14cm}{\emph{Note:} This treatment letter highlights the deadline for MSME taxpayers to submit their tax return (SPT) and emphasizes the importance of taxpayer contributions in supporting government programs, particularly those addressing the impact of the Covid-19 pandemic. The letter is issued in accordance with the official correspondence procedures of the tax office and is delivered to the taxpayer’s registered address listed in the DGT database. To verify the letter’s authenticity, taxpayers can scan the QR code linked to the electronic signature. This will display a digital copy of the letter within the Ministry of Finance’s official correspondence system.}}
\end{center}

\newpage

\begin{center}
\small{\textsc{Treatment 3: Flyer}}\\[.1cm]
\begin{textblock*}{2cm}(4.2cm, 9cm) 
	\rotatebox{90}{\footnotesize Original}
\end{textblock*}

\begin{textblock*}{2cm}(4.2cm, 17cm) 
	\rotatebox{90}{\footnotesize Translation}
\end{textblock*}
\includegraphics[width=12cm]{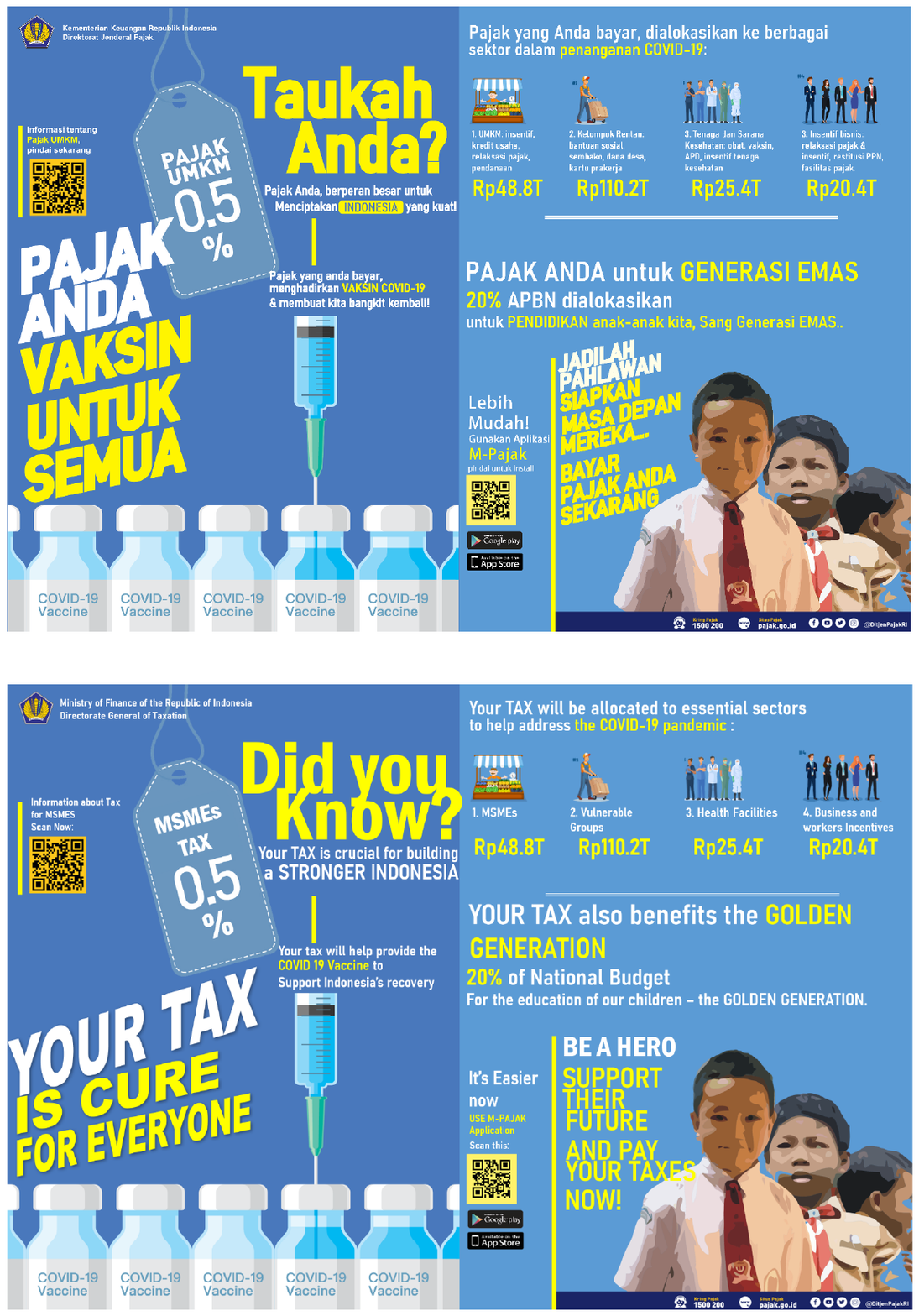}
\end{center}

\scriptsize
\centerline{\parbox{16cm}{\emph{Note:} This guide highlights the importance and impact of paying taxes, particularly for MSMEs:
\textbf{Left Panel:} Did you know? Your taxes play a crucial role in building a strong Indonesia. The taxes you pay have funded Covid-19 vaccines and helped us rise again. With an MSME tax rate of 0.5\%, your contribution acts as a ``vaccine'' for all.
\textbf{Right Panel:} The taxes you pay are allocated across various sectors to address the challenges of Covid-19. This includes IDR 4.8 trillion for MSME incentives, business loans, and tax relaxations; IDR 110.2 trillion for vulnerable groups, social assistance, food supplies, village funds, and pre-employment cards; IDR 25.4 trillion for health workers and infrastructure, medicines, vaccines, personal protective equipment, and health worker incentives; and IDR 20.4 trillion for business incentives, tax relaxations, VAT refunds, and tax facilities.
Taxes also secure the future for the ``golden generation'', with 20\% of the state budget (APBN) allocated for children’s education. Be a hero and prepare their future~-- pay your taxes now. For convenience, use the M-Pajak application. Scan the QR code to download it from the App Store or Google Play.}}

\newpage
\begin{center}
\small{\textsc{M-Pajak Mobile Apps}}\\
\includegraphics[width=16cm]{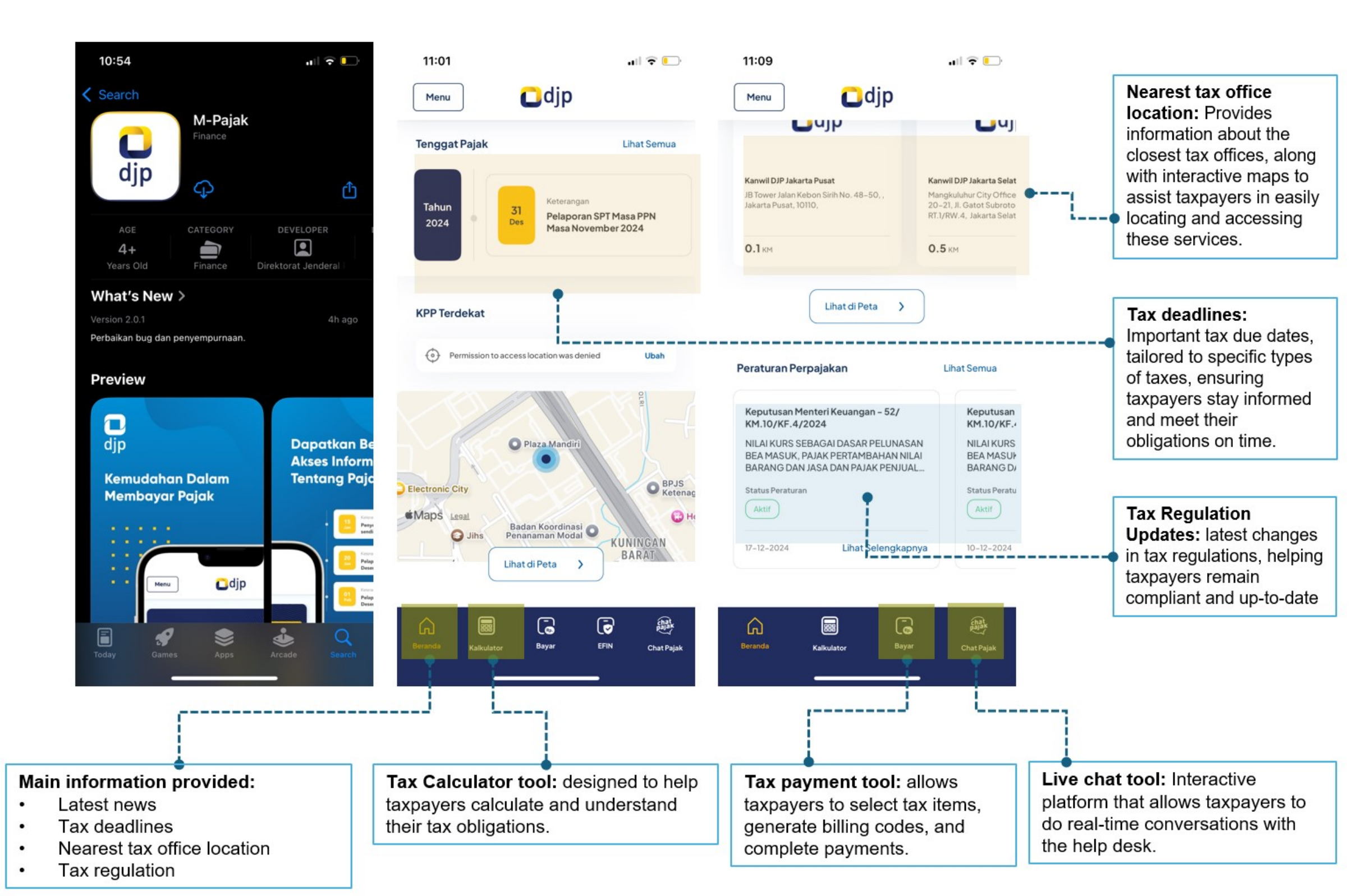}
\\[2mm]
\scriptsize
\centerline{\parbox{16cm}{\emph{Note:} M-Pajak is an official IOS/Android-based application developed by the DGT, aims to enhance the efficiency of managing taxpayers' rights and obligations by offering a comprehensive suite of tax-related services. The platform facilitates easy access to tax regulations and updates while providing timely reminders for critical tax deadlines. One of the feature of M-Pajak is its document verification and validation system, which employs QR codes to ensure authenticity. Additionally, the app offers detailed taxpayer profiles, including personal, contact, and tax-specific information, as well as information on nearby tax office locations. 
For taxpayers requiring assistance with forgotten Electronic Filing Identification Numbers (EFIN), the application includes a recovery feature that eliminates the necessity of visiting a tax office. Users can also utiliz
e the tax calculator to perform simulations of various tax computations with accuracy and speed. Furthermore, the app provides live chat support with help-desk agents and a tax office locator based on the user’s geographic location. Another key feature of M-Pajak is the generation of billing codes for online tax payments through the ``Pay'' menu. Subsequently, payments can be conveniently made via banks, ATMs, or e-commerce platforms.}}
\end{center}

\end{document}